%% file: HIG-21-020_temp.tex
\pdfoutput=1
\documentclass[11pt,twoside,a4paper,cmspaper,final,collab]{cms-tdr}
\def\svnVersion{4660184}\def\svnDate{2024/12/09}\def\cmsCernNoTag{CERN-EP-2024-162}\def\cmsCernDate{\today}\def\cmsMessage{Published in the Journal of High Energy Physics as \href{http://dx.doi.org/10.1007/JHEP12(2024)035}{\doi{10.1007/JHEP12(2024)035}.}}
\begin{document}\cmsNoteHeader{HIG-21-020}

\cmsNoteHeader{HIG-21-020}

\newcommand{\msd}{m_\mathrm{SD}}
\newcommand{\fxfx}{{\textsc{FxFx}}\xspace}

\title{Measurement of boosted Higgs bosons produced via vector boson fusion or gluon fusion in the \texorpdfstring{$\PH\to\bbbar$}{H to b-bbar} decay mode using LHC proton-proton collision data at \texorpdfstring{$\sqrt{s} = 13\TeV$}{sqrt(s) = 13 TeV}}

\date{\today}

\abstract{A measurement is performed of Higgs bosons produced with high transverse momentum (\pt) via vector boson or gluon fusion in proton-proton collisions. The result is based on a data set with a center-of-mass energy of 13\TeV collected in 2016--2018 with the CMS detector at the LHC and corresponds to an integrated luminosity of 138\fbinv. The decay of a high-\pt Higgs boson to a boosted bottom quark-antiquark pair is selected using large-radius jets and employing jet substructure and heavy-flavor taggers based on machine learning techniques. Independent regions targeting the vector boson and gluon fusion mechanisms are defined based on the topology of two quark-initiated jets with large pseudorapidity separation.  The signal strengths for both processes are extracted simultaneously by performing a maximum likelihood fit to data in the large-radius jet mass distribution.  The observed signal strengths relative to the standard model expectation are $4.9^{+1.9}_{-1.6}$ and $1.6^{+1.7}_{-1.5}$ for the vector boson and gluon fusion mechanisms, respectively. A differential cross section measurement is also reported in the simplified template cross section framework.}

\hypersetup{
  pdfauthor={CMS Collaboration},
  pdftitle={Measurement of boosted Higgs bosons produced via vector boson fusion or gluon fusion in the H to b-bbar decay mode using LHC proton-proton collision data at sqrt(s) = 13 TeV},
  pdfsubject={CMS},
  pdfkeywords={CMS, Higgs, vector boson fusion}
}

\maketitle

\section{Introduction}
\label{intro}

The observation at the CERN LHC of a Higgs boson (\PH) consistent with the standard model (SM) expectation and the subsequent measurements of its properties~\cite{Aad_2012,Chatrchyan_2012,cms2013,CMS:2022dwd, ATLAS:2022vkf,201388} have advanced the understanding of electroweak (EW) symmetry breaking and the origin of the mass of fundamental particles~\cite{Salam:1968rm,Glashow:1961tr,Weinberg:1967tq,PhysRevLett.13.321,Higgs:1964ia,PhysRevLett.13.508,PhysRev.145.1156,PhysRevLett.13.585}.
Higgs boson production at high momentum transfer can be a sensitive probe of beyond the SM (BSM) physics at high energy scales, and measurements in this regime provide important input for effective field theory interpretations of \PH interactions~\cite{Grojean_2014,Schlaffer_2014,Dawson_2015,Grazzini_2017,Maltoni_2014,Degrande_2017}. Thus, studying Higgs bosons at high transverse momentum ($\pt$) has become an integral part of the physics program at the LHC.

The Higgs boson decay to a bottom quark-antiquark pair ($\bbbar$) has the highest branching fraction of any \PH decay mode in the SM~\cite{Badger:2016bpw}.
When the Higgs boson is produced with high momentum, the $\PH\to\bbbar$ decay products are merged into a single large-radius jet. This distinctive boosted jet topology can be identified by its substructure and heavy flavor properties, and as a result the $\PH\to\bbbar$ decay provides an important channel for the exploration of Higgs boson production at high $\pt$.

\begin{figure}[htbp]
  \centering
  \includegraphics[width=0.95\textwidth]{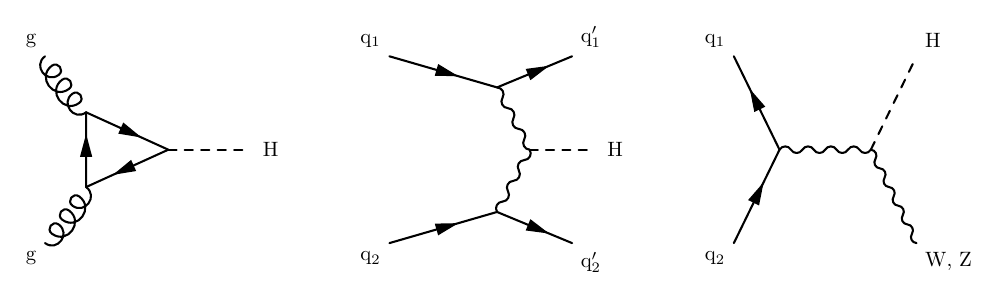}
  \caption{Lowest order Feynman diagrams of the Higgs boson production modes with highest cross section in $13\TeV$ proton-proton collisions, from left to right: gluon fusion, vector boson fusion, and vector boson associated production.
  }
  \label{fig:feynman}
\end{figure}

Leading order (LO) Feynman diagrams of the three \PH production mechanisms with highest cross section in proton-proton ($\Pp\Pp$) collisions are shown in Fig.~\ref{fig:feynman}: gluon fusion (ggF), vector boson fusion (VBF), and vector boson associated production ($\PV\PH$). A small contribution to the \PH cross section also arises from associated production with a top quark-antiquark pair ($\ttbar\PH$).
The dominant SM contribution comes from ggF, which contributes 87\% of the \PH cross section at $\sqrt{s}=13\TeV$ when considering the full range of Higgs boson \pt.
However, the relative contribution from ggF is expected to decrease with the \pt of the Higgs boson, contributing only 50 (30)\% of the cumulative \PH cross section for $\pt>450$ (1200)\GeV~\cite{becker2021precise}.
While the relative contribution from $\PV\PH$ increases as a function of Higgs boson \pt, that of VBF has minimal \pt dependence.
Each of these production mechanisms provides a different probe of \PH interactions, and precise measurement of each one is necessary to investigate all possible manifestations of BSM physics in the Higgs sector.

Existing searches in the boosted $\PH\to\bbbar$ channel from the ATLAS and CMS experiments have focused on inclusive Higgs boson production~\cite{Sirunyan_2018,Aad_2022,Sirunyan_2020} or associated production with a vector boson~\cite{ATLAS_VHlep,ATLAS_VHhad,HIG-20-001}.
This work extends, for the first time, high-$\pt$ Higgs boson measurements in the $\PH\to\bbbar$ channel to VBF production.  The analysis is performed using pp collision data collected with the CMS detector at the LHC in 2016--2018, corresponding to an integrated luminosity of 138\fbinv~\cite{cmsLumi16,CMS-PAS-LUM-17-004,CMS-PAS-LUM-18-002}. The VBF process, which is sensitive to \PH couplings to vector bosons, and the ggF process, which is primarily sensitive to \PH couplings to top quarks and gluons, are measured simultaneously.  In order to disentangle the two processes, the analysis is performed in two categories. The first category is enriched in VBF events and targets the characteristic topology of two quark-initiated jets with high pseudorapidity ($\eta$). The second category is enriched in ggF events.

In both categories, the $\PH\to\bbbar$ decay mode is isolated from quark- and gluon-initiated jets by a multivariate large-radius jet classifier known as the DeepDoubleB (DDB) tagger~\cite{CMS-DP-2022-041}. Updates with respect to the previous version of the DDB tagger~\cite{CMS-DP-2018-046,Sirunyan:2017ezt} yield a gain in expected signal significance of about a factor of 2 compared to the previous CMS $\PH\to\bbbar$ search~\cite{Sirunyan_2020}.
As a result, the analysis reported here provides not only the first measurement of \PH production via VBF at high $\pt$, but also the most precise measurement of ggF Higgs boson production in this regime, superseding the measurement in Ref.~\cite{Sirunyan_2020}.

\section{The CMS detector}

The CMS apparatus~\cite{cmsDetector} is a multipurpose, nearly hermetic detector, designed to trigger on~\cite{cmsTrigger,cmsTrigger2} and identify electrons, muons, photons, and hadrons~\cite{cmsElectronReco,cmsMuon,cmsTracking}. Information about particles traversing the detector is provided by the all-silicon inner tracker, and by the crystal electromagnetic and brass-scintillator hadron calorimeters, all of which operate inside a 3.8\unit{T} superconducting solenoid, and by the gas-ionization muon detectors embedded in the flux-return yoke outside the solenoid.

Events of interest are selected using a two-tiered trigger system. The first level, composed of custom hardware processors, uses information from the calorimeters and muon detectors to select events at a rate of around 100\unit{kHz} within a fixed latency of about 4\mus~\cite{cmsTrigger}. The second level, known as the high-level trigger, consists of a farm of processors running a version of the full event reconstruction software optimized for fast processing, and reduces the event rate to around 1\unit{kHz} before data storage~\cite{cmsTrigger2}.

A more detailed description of the CMS detector, together with a definition of the coordinate system used and the relevant kinematic variables, can be found in Ref.~\cite{cmsDetector}.

The dataset used in this analysis is divided into four data-taking periods to account for the 2017 upgrade of the CMS tracking detector, differences in accelerator and detector conditions, and updates to reconstruction algorithms and calibrations. These data-taking periods correspond to early 2016, late 2016, 2017, and 2018. Data from 2016 is split into an early and a late data-taking period due to changes made mid-year.

\section{Simulated samples}
\label{sec:samples}

Simulated samples of signal and background events are produced using various Monte Carlo (MC) event generators, with the CMS detector response modeled by \GEANTfour~\cite{GEANT4}. Independent samples are generated for each of the four data-taking periods using identical generator configurations, but accounting for changes in the accelerator and detector running conditions.

The background from jets produced via the strong interaction, referred to as quantum chromodynamics (QCD) multijet events, is modeled at LO accuracy using the \MGvATNLO~ 2.6.5 generator~\cite{Alwall_2014} with up to four partons from the matrix element calculation.

The $\PW$+jets and $\PZ$+jets processes are modeled at LO accuracy using the \MGvATNLO generator.
These samples include decays of the bosons to all flavors of quarks and up to three (four) extra partons at the matrix element level for $\PW$+jets ($\PZ$+jets).
Jets from the matrix element calculations and parton shower description are matched using the MLM prescription~\cite{Alwall_2007}.  Correction factors are applied to the $\PW$+jets and $\PZ$+jets samples to match the generator-level \pt distributions with those predicted by the highest available order in the perturbative expansion.
The QCD next-to-LO (NLO) corrections are derived using \MGvATNLO, simulating \PW and \PZ boson production with up to two additional partons and \fxfx matching to the parton shower~\cite{Frederix:2012ps}.
The EW NLO corrections are taken from theoretical calculations in Refs.~\cite{kallweit2015nlo,Kallweit_2016,kallweit2015nlo2,Lindert_2017}.

Electroweak production of $\PW$ and $\PZ$ bosons is modeled using \MGvATNLO at LO accuracy. Diboson processes are modeled at LO accuracy with \PYTHIA~8.226~\cite{Sjostrand:2014zea}, and the total cross sections are corrected to next-to-NLO (NNLO) accuracy with the \MCFM~7.0 program~\cite{Campbell:2010ff}.  Top quark-antiquark (\ttbar) and single top quark production are modeled at NLO in QCD using \POWHEG~2.0~\cite{Frixione:2007vw,Alioli:2010xd,Frixione:2007nw,Frederix:2012dh,Re:2010bp}.

The ggF Higgs boson production process is simulated using the HJMINLO event generator~\cite{Monni:2019whf,Monni:2020nks} with the Higgs boson mass set to $125\GeV$.  Finite top quark mass effects~\cite{Neumann_2018} are included following the recommendation in Ref.~\cite{becker2021precise}.
The \POWHEG generator~\cite{Nason:2009ai,Luisoni:2013kna,Hartanto:2015uka} is used to model Higgs boson production via VBF, $\PV\PH$, and $\ttbar\PH$ at NLO accuracy.
The VBF sample is reweighted to account for NNLO corrections to the $\pt$ spectrum and N$^3$LO corrections to the inclusive cross section~\cite{Cacciari:2015jma,Dreyer:2016oyx}.
These corrections have a negligible effect on the yield for this process for events with Higgs boson $\pt>450\GeV$. EW corrections are also applied to the VBF, $\PV\PH$, and $\ttbar\PH$ processes~\cite{becker2021precise}.

For parton showering and hadronization, all generated samples are interfaced with \PYTHIA~8.230, with the parameters for the underlying event description set via the CP5 tune~\cite{cmsTune}.  The diboson process exceptionally uses \PYTHIA~8.226 for both the matrix element and parton shower evolution.  The parton distribution function (PDF) set {NNPDF3.1}~\cite{Ball_2017} at NNLO accuracy is used for all processes.

\section{Event reconstruction and selection}

The particle-flow (PF) algorithm~\cite{cmsPF} aims to reconstruct and identify each individual particle in an event using an optimized combination of information from the various elements of the CMS detector.  The primary vertex is taken to be the vertex corresponding to the hardest scattering in the event, evaluated using tracking information alone, as described in Ref.~\cite{cmsTracking}.

Jets are reconstructed by clustering PF candidates using the anti-\kt algorithm~\cite{Cacciari_2008} as implemented in the \FASTJET software package~\cite{Cacciari_2012}. Small-radius jets are clustered with a distance parameter of 0.4. Large-radius jets, which are used to capture the decay products of heavy boosted objects, use a distance parameter of 0.8~\cite{CMS-PAS-JME-16-003}.

Jet momentum is taken to be the vectorial sum of all PF candidate momenta in the jet. Additional pp interactions within the same or nearby bunch crossings, known as pileup, can increase the apparent jet momentum by contributing additional tracks and calorimetric energy deposits. To mitigate this effect
for small-radius jets, tracks originating from pileup vertices are discarded and an offset correction is applied to correct for remaining contributions~\cite{CMS:2020ebo}.
For large-radius jets, the pileup per particle identification algorithm~\cite{PUPPI} is used to weight the PF candidates prior to jet clustering. Jet energy corrections are derived from simulation to bring the measured response of jets to that of particle level jets on average~\cite{CMS:2016lmd}. Additional selection criteria are applied to remove jets that are likely to be dominated by instrumental effects or reconstruction failures~\cite{CMS:2020ebo,CMS-PAS-JME-16-003}.

To isolate the \PH signal, a large-radius jet with high \pt is required.  Events are selected by a combination of triggers, with each trigger imposing a minimum threshold on either the $\pt$ of a large-radius jet or the event $\HT$, the scalar $\pt$ sum of all small-radius jets in the event with $\abs{\eta}<3$. For large-radius jets used in the trigger selection, a minimum trimmed jet mass~\cite{jetTrimming} is also required. The trigger selection is about 90\% efficient with respect to the offline selection for events containing large-radius jets with \pt from $450$--$500\GeV$ and $\abs{\eta}<2.5$, and is fully efficient for $\pt>500\GeV$ and $\abs{\eta}<2.5$.  In order to ensure high trigger efficiency, each event is required to contain at least one large-radius jet with $\pt > 450\GeV$ and $\abs{\eta} < 2.5$. Large-radius jets are required to pass the tight jet identification requirement~\cite{CMS-PAS-JME-16-003} and be separated from all isolated photons or charged leptons by a distance of $\Delta R = \sqrt{\smash[b]{\Delta\eta\,^2 + \Delta\phi\,^2}} > 0.8$, where $\Delta\eta$ and $\Delta\phi$ are the differences in pseudorapidity and azimuthal angle, respectively.

For large-radius jets originating from the decay of a massive boson (\PW, \PZ, \PH), contamination from underlying event and pileup can lead to an overestimate of the jet invariant mass and degraded jet mass resolution. The soft drop algorithm~\cite{softdrop}, with parameters $\beta=0$ and $z_\text{cut}=0.1$, is applied to mitigate these effects by removing PF candidates consistent with soft and wide-angle radiation from the jet. For jets originating from the decay of a massive boson, the soft drop mass ($\msd$) peaks at the boson mass, while for quark- and gluon-initiated jets, $\msd$ has a smoothly falling spectrum.

In order to avoid regions where non-perturbative effects lead to discrepancies in generator predictions~\cite{ATLASrho}, large-radius jets are required to have $\rho>-6.0$, where $\rho=2\ln(\msd/\pt)$. Large-radius jets with $\rho>-2.1$ are also vetoed, since above this threshold the jet constituents are no longer fully contained in a cone of $R=0.8$. This $\rho$ selection is fully efficient for \PH signal events.

The multivariate DDB tagger discriminant is used to determine whether a large-radius jet originates from a heavy particle decaying to a $\bbbar$ pair~\cite{CMS-DP-2022-041}. The DDB tagger is based on a deep neural network that is trained to distinguish large-radius jets containing scalar $\PX\to\bbbar$ decays from jets originating from light quarks or gluons.
The DDB tagger is trained on jet features, such as the number of PF
candidates and low-level quantities describing secondary vertices within the jet.
The addition of low-level features leads to a large improvement in performance relative to the previous version of the tagger~\cite{CMS-DP-2018-046,Sirunyan:2017ezt}: for a QCD jet mistag rate of 1\%, the signal efficiency of the updated DDB tagger is about 75\%, representing an improvement of about 50\% with respect to the previous version.

\begin{figure}[htbp]
  \centering
  \includegraphics[width=0.45\textwidth]{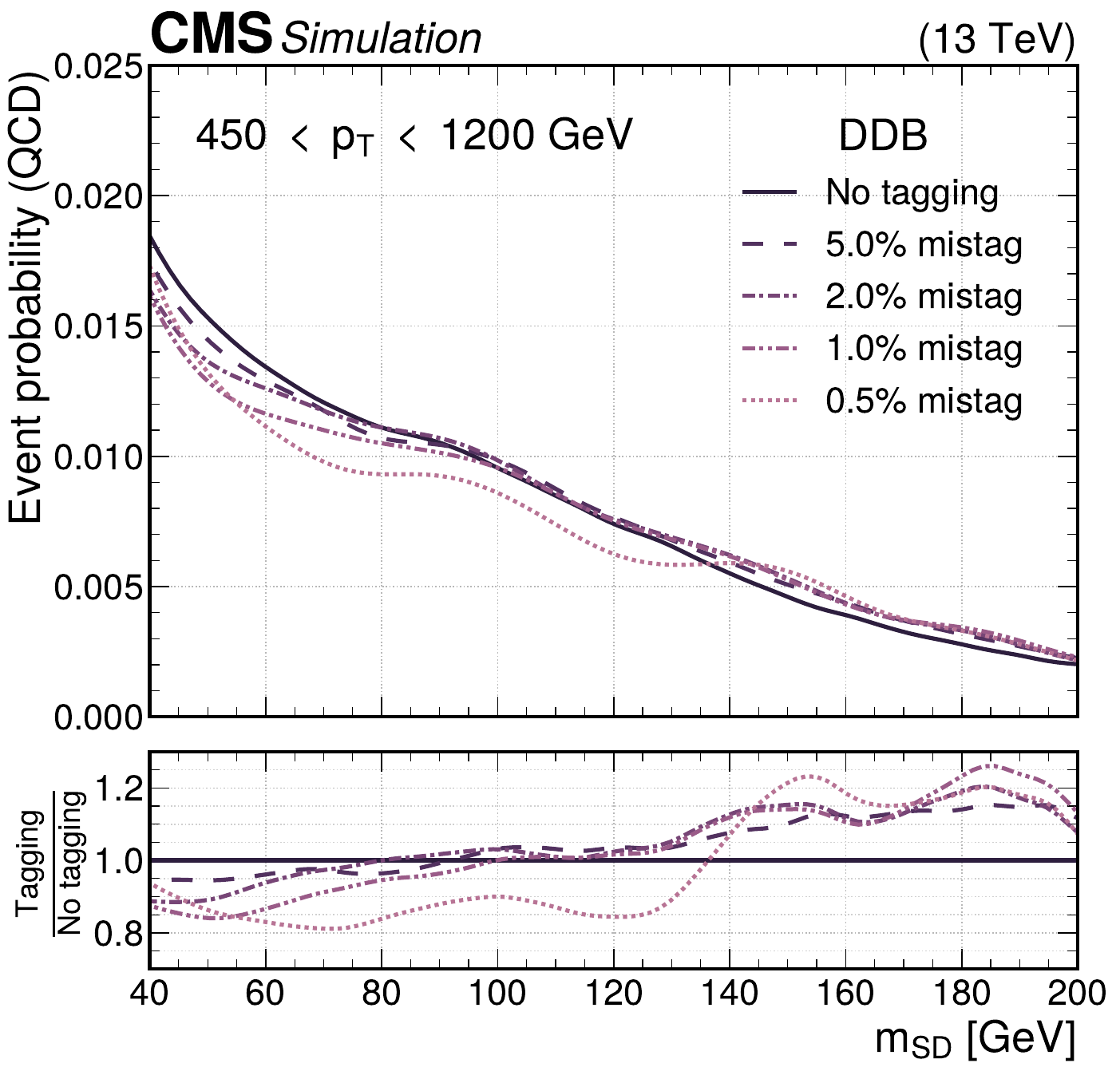}
  \caption{Soft drop mass distribution in simulated QCD events after applying the DDB selection at different working points. The distributions are obtained from simulated QCD events, smoothed using Gaussian kernel density estimation, and normalized to unit area. The lower panel shows the ratio to the inclusive distribution.}
  \label{sel-ddb}
\end{figure}

In order to ensure that the DDB tagger selection does not impact the shape of the background $\msd$ distribution, the DDB network is trained on a signal sample composed of simulated decays of bosons with masses ranging from 20 to 200\GeV. The success of this decorrelation method is demonstrated in Fig.~\ref{sel-ddb}, which shows the shape of the $\msd$ distribution for simulated QCD jets with $450 < \pt < 1200\GeV$ after different DDB selection criteria have been applied. The shape of the $\msd$ distribution is shown to be mostly independent of the DDB threshold value.

The large-radius jet with the highest DDB score is identified as the Higgs boson candidate. This criterion selects a \PH candidate jet within $\Delta R<0.8$ of the generated particle in more than 90\% of simulated $\PH\to\bbbar$ events.  Events in which the selected candidate jet does not correspond to the \PH decay are excluded from the signal yield, and their contribution is accounted for in the estimation of the QCD multijet background. 

Events containing \PH candidate jets with high DDB tagger score enter the ``DDB pass'' region.  The definition of the DDB pass region is chosen to maximize the expected sensitivity to VBF \PH production, and the DDB tagger threshold corresponds to a signal efficiency of about 40\% and a QCD multijet mistag rate of 0.5\%.  Events containing \PH candidate jets below this DDB threshold enter the ``DDB fail'' region, which is later used to constrain the normalization and shape of the QCD multijet background.

Higgs boson candidate jets are required to be consistent with the two-prong substructure of a $\PH\to\bbbar$ decay, reducing the contribution from $\ttbar$, single top, and QCD backgrounds. The substructure selection is based on the $N_2^1$ variable, a ratio of three-point to two-point energy correlation functions proposed in Ref.~\cite{Moult_2016}. In order to minimize the dependence of the substructure selection on the $\pt$ and $\msd$ of the \PH candidate, a variant of $N_2^1$ is defined using the designed decorrelated tagger (DDT) technique~\cite{Dolen_2016}. The value of $N_2^1$ that results in a background efficiency $\epsilon$ is determined as a function of jet $\pt$ and $\rho$ and denoted $N_\epsilon(\pt,\rho)$. A decorrelated substructure variable, which has a background efficiency that is independent of jet $\pt$ and $\rho$, can be obtained by subtracting $N_\epsilon(\pt,\rho)$ from the nominal substructure variable. The following transformation defines the decorrelated substructure variable used in this analysis:
\begin{equation}
  N_2^{1,\mathrm{DDT}} = N_2^1 - N_{\epsilon}(\pt,\rho).
\end{equation}

A background efficiency of $\epsilon=0.26$ is chosen to maximize the total \PH signal sensitivity, and $N_{0.26}(\pt,\rho)$ is derived from the simulated multijet distribution of $N_2^1$ separately in each data-taking period. Selected events are required to have $N_2^{1,\text{DDT}}<0$, which retains about 50\% of \PH signal events and (by construction) rejects 74\% of expected QCD events. The calibration of this substructure requirement and the associated uncertainties are described in Section~\ref{sec:vjets}.

Small-radius jets are required to have $\pt > 30\GeV$ and $\abs{\eta} < 5.0$. These jets must be separated from the \PH candidate jet by a distance of $\Delta R > 0.8$. The \textsc{DeepJet} algorithm~\cite{Bols:2020bkb,CMS-DP-2018-058} is used to identify small-radius jets within the volume of the silicon tracking detectors ($\abs{\eta} < 2.5$) that originate from \PQb quarks. In order to reduce contamination from $\ttbar$ background, events are vetoed if any of the four highest-$\pt$ small-radius jets in the hemisphere opposite the \PH candidate is \PQb tagged.

The missing transverse momentum vector \ptvecmiss is computed as the negative vector sum of the transverse momenta of all the PF candidates in an event, with modifications to account for corrections to the energy scale of the reconstructed jets in the event~\cite{CMS:2019ctu}. The magnitude of \ptvecmiss is denoted \ptmiss and required to be less than 140\GeV.

Events containing charged leptons are vetoed.  For purposes of this veto, electrons must have $\pt>10\GeV$ and $\abs{\eta}<2.5$ and pass the loose identification criteria outlined in Ref.~\cite{cmsElectronReco}. Muons must satisfy the loose identification criteria described in Ref.~\cite{cmsMuon} and have $\pt>10\GeV$ and $\abs{\eta}<2.4$. For electrons (muons), the isolation variable corresponds to the pileup-corrected sum $\pt$ of charged hadrons and neutral particles in a cone of $R=0.3$ (0.4) around the lepton divided by the lepton $\pt$. The isolation is required to be less than 0.15 and 0.25 for electrons and muons, respectively. Hadronically decaying tau leptons ($\tau_\text{h}$) are reconstructed from jets using the hadrons-plus-strips algorithm~\cite{cmsTau}.  To distinguish genuine $\tau_\text{h}$ decays from jets originating from the hadronization of quarks or gluons, as well as from electrons or muons, the \textsc{DeepTau} algorithm is used~\cite{Tumasyan_2022}. For the purposes of a lepton veto, $\tau_\text{h}$ must have $\pt> 20\GeV$ and $\abs{\eta}< 2.3$.

\begin{figure}[htbp]
  \centering
  \includegraphics[width=0.7\textwidth]{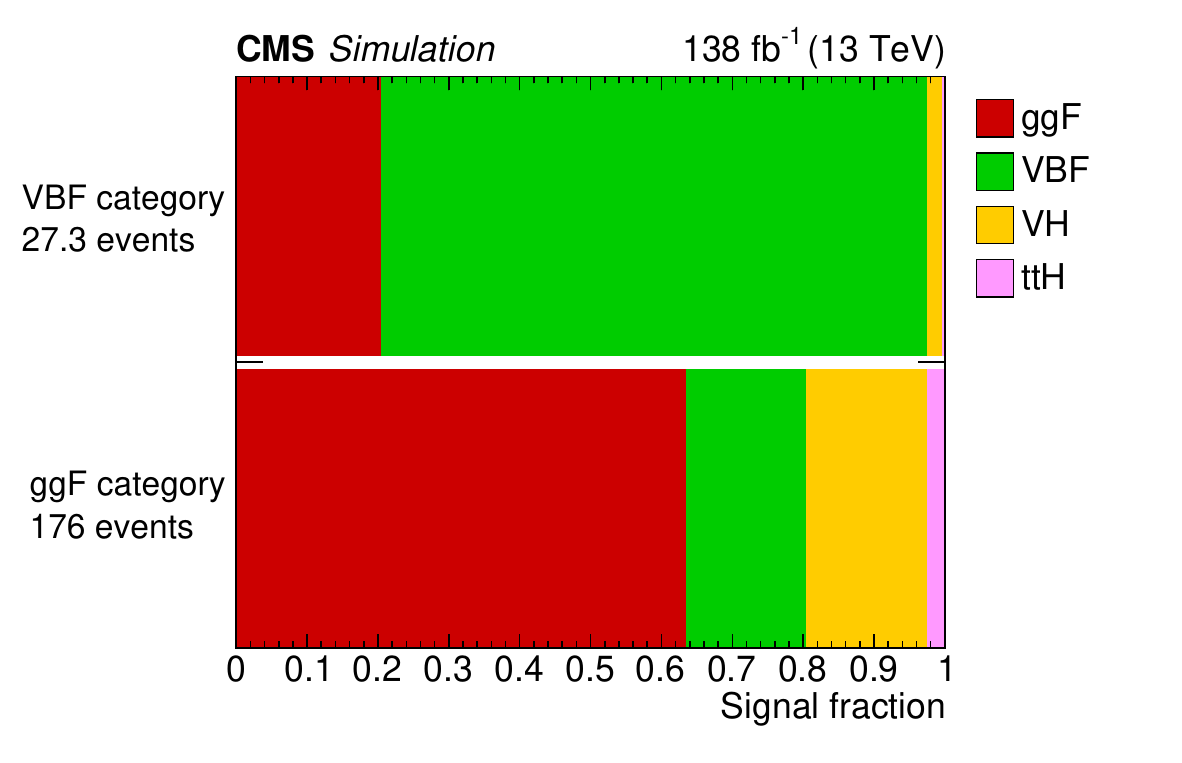}
  \caption{Simulated relative contributions of the four leading \PH production processes to the total \PH signal yield in the VBF and ggF categories, shown in the DDB pass region. Small contributions from the $\PV\PH$ and $\ttbar\PH$ processes are also included. The total number of predicted \PH events in the 138\fbinv dataset are 27.3 and 176 in the VBF and ggF categories, respectively.}
  \label{purity}
\end{figure}

In order to measure the VBF and ggF production mechanisms simultaneously, the selected events (both DDB pass and fail) are partitioned into two categories. The VBF category selection is optimized for sensitivity to VBF production by targeting the characteristic forward (high $\abs{\eta}$) quark-initiated jets. Events entering this category are required to contain two small-radius jets in addition to the \PH candidate. The small-radius jets are required to have a pseudorapidity separation of $\abs{\Delta \eta}>3.5$ and invariant mass of $m_\text{jj}>1\TeV$, where the selection criteria are chosen to maximize the sensitivity to VBF production. If an event contains more than two small-radius jets, the highest and second highest in $\pt$ are used to construct $\Delta \eta$ and $m_\text{jj}$. Events containing fewer than two small-radius jets, or where $\abs{\Delta \eta}<3.5$ or $m_\text{jj}<1\TeV$, fall into the ggF category. Figure~\ref{purity} shows the predicted relative contribution of each production mechanism to the total Higgs boson signal in the VBF and ggF categories.

The VBF category is divided into two bins in $m_\text{jj}$ with different signal purity: $1<m_\text{jj}<2\TeV$ and $m_\text{jj}>2\TeV$. When combined statistically, this yields about 20\% improvement in the expected significance of a VBF signal compared to a single bin.
Similarly, the ggF category is divided into six bins in the \PH candidate jet $\pt$ with boundaries 450, 500, 550, 600, 675, 800, and 1200\GeV. The upper \pt bound of 1200\GeV does not have a significant impact on the sensitivity and excludes a region where the QCD multijet background is difficult to model. The subdivision into $\pt$ bins improves the expected ggF significance by about 20\%.

\section{Background estimation}

Because of its large cross section, QCD multijet production is the dominant background in the combined DDB pass and fail regions. Significant resonant backgrounds arise from $\PW$+jets and $\PZ$+jets processes, where the two-prong decay of a high-$\pt$ vector boson is mistagged as a \PH candidate. Electroweak \PW and \PZ boson production contribute in particular to the background in the VBF category because of the presence of two forward quark-initiated jets. For events in which a high-$\pt$ top quark is produced, it is possible that a large-radius jet captures only two of the three prongs of the decay and passes the $N_2^{1,\text{DDT}}$ selection. The presence of a bottom quark in these decays makes them more likely to pass the DDB requirement. In this way, top quark production contributes to the nonresonant component of the background.  This section describes the estimate of the background contribution from each of these processes.

Additional small background contributions from diboson, $\PV\PH$, and $\ttbar\PH$ processes (each expected to be $<1\%$ of the total background) are estimated from simulation. Both $\PV\PH$ and $\ttbar\PH$ are assumed to occur at SM rates.

\subsection{QCD background}
\label{sec:qcdbkg}

The soft drop jet mass in QCD multijet events follows a steeply falling distribution.
The predicted shape of the QCD background in the DDB pass region is derived using data in the background-enriched DDB fail region.
The full prediction of the QCD background yield in the DDB pass region takes the form
\begin{equation}
  N_\text{P}^i =  N_\text{F}^{\text{data},i} \; R_\text{P/F}^{\text{MC},i} \; T_\text{P/F}(\pt^i,\rho^i) \; T_\text{res}(\pt^i,\rho^i),
\end{equation}
where bin $i$ corresponds to $\pt^i$ and $\rho^i$. $N_\text{F}^{\text{data},i}$ is the number of data events minus the number of predicted non-QCD background events in bin $i$ in the DDB fail region. The ratio $R_\text{P/F}^{\text{MC},i}$ is the number of simulated QCD events in the DDB pass region divided by that in the DDB fail region.

Two transfer factors, $T_\text{P/F}$ and $T_\text{res}$, account for shape differences in the jet \pt and $\msd$ distributions between the DDB pass and fail regions.
In the ggF category, the functional form of each transfer factor is taken to be a two-dimensional Bernstein polynomial in the \PH candidate jet $\pt$ and $\rho$:
\begin{equation}
  T_\text{P/F}(\pt,\rho) = \sum^{n_\rho}_{k=0}\sum^{n_{\pt}}_{l=0}a_{k,l}\left[ b_{k,n_\rho}(\rho) b_{l,n_{\pt}}(\pt)\right],
\end{equation}
where $n_{\rho}$ and $n_{\pt}$ are the degrees of the polynomial in $\rho$ and $\pt$, respectively, $a_{k,l}$ are fitted coefficients, and $b_{\nu,n}$ are one-dimensional Bernstein polynomials:
\begin{equation}
  b_{\nu,n}(x) = \binom{n}{\nu} x^\nu(1-x)^{n-\nu}.
\end{equation}
The variables $\pt$ and $\rho$ are affine transformed to lie within the domain of the Bernstein basis. 
Because the VBF category is not binned in \pt, the transfer factors in this category are parameterized only in $\rho$, and a sum of one-dimensional Bernstein polynomials is used.

The first transfer factor, $T_\text{P/F}$, accounts for possible \pt- or $\msd$-dependent effects due to the DDB selection by parameterizing the ratio of the QCD event yield in the DDB pass and fail regions. The coefficients $a_{k,l}$ of $T_\text{P/F}$ are derived with a dedicated fit to the $\msd$ distribution in QCD MC simulation.  Discrepancies in the tagger performance between data and simulation are accounted for with a second transfer factor, $T_\text{res}$.
The coefficients of $T_\text{res}$ are obtained in the final simultaneous fit to data described in Section~\ref{sec:results}.
Independent transfer factors $T_\text{P/F}$ and $T_\text{res}$ are defined for each data-taking period, and in the ggF category and each of the two $m_\text{jj}$ bins in the VBF category.

For each transfer factor, the optimal number of free parameters is determined by a Fisher F-test~\cite{ftest}. As a first step, a low-order polynomial of $n_1$ parameters is taken as the baseline function. An alternative function with $n_2 > n_1$ parameters is tested against the baseline. In the case where the baseline function has $n_\rho = n_{\pt}$, the alternative function is determined by increasing the polynomial order in $\rho$. The alternative function is adopted as the new baseline if it provides a significantly better goodness-of-fit (F-statistic with $p$-value~\cite{Gross_2010} less than 0.05).
The polynomial order is optimized independently for $T_\text{P/F}$ and $T_\text{res}$, separately in each data-taking period and in the ggF category and each of the two $m_\text{jj}$ bins in the VBF category.

In the ggF category, the optimal polynomial order for $T_\text{P/F}$ is determined to be $(n_{\pt}, n_\rho) = (1,1)$ for early 2016 and $(n_{\pt}, n_\rho) = (0,1)$ for late 2016, 2017, and 2018. The optimal polynomial order for $T_\text{res}$ is found to be $(n_{\pt}, n_\rho) = (0,2)$ for all data-taking periods. In the VBF category, the optimal polynomial order for both transfer factors is found to be $n_\rho=0$ for all data-taking periods in both $m_\text{jj}$ bins.

Tests with alternative functional forms for the transfer factors show that the final result has negligibly small dependence on the choice of the parameterization.

\subsection{W+jets and Z+jets backgrounds}
\label{sec:vjets}

The $\PW$+jets and $\PZ$+jets backgrounds are modeled using simulation, with normalizations and kinematic distributions corrected for NLO QCD and EW effects, as described in Section~\ref{sec:samples}. A scale factor ($f_\text{sub}$) is applied to the $\PW$+jets and $\PZ$+jets processes to correct for the efficiency of the two-prong substructure selection ($N_2^{1,\text{DDT}}<0$). A dedicated \PW tag control region (CR) is used to simultaneously derive $f_\text{sub}$, a correction factor on the jet mass resolution ($f_\sigma$), and a jet mass shift ($\delta_m$). These corrections are applied to the predictions from simulation for $\PW$+jets and $\PZ$+jets, as well as all \PH signal processes.

Events entering the \PW tag CR must pass a trigger requiring a muon with $\pt>50\GeV$. Events must contain a muon passing the tight identification requirements defined in Ref.~\cite{cmsMuon} and have $\ptmiss> 40\GeV$. The $\pt$ of the muon plus $\ptmiss$ system, which forms a \PW boson proxy, must be greater than $200\GeV$. Finally, at least one \PQb-tagged small-radius jet is required. This region contains a high fraction of $\ttbar$ events, and the substructure variable requirements select primarily top quark decays to boosted \PW bosons.

Two template $\msd$ distributions are constructed from simulation in the \PW tag CR: events that contain a \PW boson, and events that do not. Variations in the jet mass scale and jet mass resolution are calculated for the template containing a \PW boson using linear morphing~\cite{Conway:1333496}.
The desired corrections are then extracted from a template fit to data in the  \PW tag CR. This fit is performed simultaneously in the regions passing and failing the two-prong substructure selection ($N_2^{1,\text{DDT}}<0$ and $N_2^{1,\text{DDT}}>0$, respectively).

The fitted values of $f_\text{sub}$, $f_\sigma$, and $\delta_m$ are shown in Table~\ref{sys-tab-sf} with their uncertainties.  These corrections are measured independently for each data-taking period in order to account for the 2017 upgrade of the CMS tracking detector, differences in pileup and detector conditions, and updates to reconstruction algorithms and calibrations.  In the fit to the signal regions in data, the uncertainties on these corrections are treated as systematic uncertainties.

\begin{table}[!htp]
  \begin{center}
    \topcaption{Summary of corrections for the jet substructure selection (scale factor $f_\text{sub}$), jet mass resolution (scale factor $f_\sigma$), and jet mass scale (shift $\delta_m$) for different data-taking periods.}
    \begin{tabular}{lllll}
      {}         & $f_\text{sub}$ & $f_\sigma$    & $\delta_m$ [\MeVns{}]   \\ \hline
      Early 2016 & $0.99\pm0.16$  & $1.13\pm0.04$ & $-240\pm410$       \\
      Late 2016  & $0.82\pm0.15$  & $1.21\pm0.03$ & $+440\pm340$       \\
      2017       & $1.05\pm0.10$  & $1.09\pm0.02$ & $+470\pm180$     & \\
      2018       & $0.94\pm0.08$  & $1.07\pm0.03$ & $-990\pm250$     & \\
    \end{tabular}
    \label{sys-tab-sf}
  \end{center}
\end{table}
A nonresonant background arises from $\PW$+jets and $\PZ$+jets events in which the selected candidate jet does not correspond to the decay of the vector boson. Since such jets originate instead from QCD processes, they are accounted for in the QCD background estimation.

\subsection{Top quark background}

Top quark processes contribute a nonresonant component of a few percent to the total background.  While the shape of the $\msd$ distributions in $\ttbar$ and single top quark processes are estimated from simulation, both the overall normalization and the DDB tagger efficiency are constrained using data. This is accomplished by including a single-muon CR, which closely mimics the signal selection described above, but selects a semileptonic decay of the $\ttbar$ pair rather than all-hadronic.

Events entering the single-muon CR must pass a trigger requiring a muon with $\pt>50\GeV$. Following trigger selection, events must contain a muon with $\pt >55\GeV$ and $\abs{\eta} < 2.1$. The muon must satisfy the loose identification requirements defined in Ref.~\cite{cmsMuon} and have isolation variable less than 0.25. Events containing electrons or $\tau_\text{h}$ are vetoed. At least one large-radius jet with $\pt > 400\GeV$,  $\abs{\eta} < 2.5$, $\msd > 40\GeV$ and $N_2^{1,\text{DDT}} < 0$ is required, and this jet must be separated from the muon by an angular distance of $\Delta \phi > 2\pi/3$.  At least one small-radius \PQb-tagged jet with $\pt > 50\GeV$ and $\abs{\eta} < 2.5$ is also required.  The resulting CR is more than 80\% pure in top quark processes. Small contributions from QCD, $\PW$+jets, and $\PZ$+jets processes are estimated from simulation.

In order to account for kinematic differences between the single-muon CR and the VBF-specific phase space,
an additional systematic uncertainty is applied to the $\ttbar$ yield in the VBF category. The magnitude of this uncertainty is derived from a comparison of the data-to-simulation scale factor obtained in the inclusive single-muon CR and in the same CR with additional requirements on the number of small-radius jets, $\Delta\eta$, and $m_\text{jj}$.

The DDB pass and fail regions of the single-muon CR are treated as single-bin counting experiments in the final fit framework.

\section{Systematic uncertainties}
\label{sec:systematics}

Several sources of experimental uncertainty arise from the reconstruction and calibration of physics objects from low-level detector responses.

A data-to-simulation scale factor is applied to the $\PH\to\bbbar$ and $\PZ\to\bbbar$ predictions to account for the difference in efficiency of the DDB tagger selection. This scale factor is included in the signal extraction fit as a constrained nuisance parameter with a nominal value of unity and a conservative uncertainty of 30\%. The magnitude of this uncertainty is based on the results of dedicated measurements in QCD multijet events with gluon to $\bbbar$ splitting topologies similar to the $\PH\to\bbbar$ signal~\cite{CMS-PAS-BTV-22-001}, but a precise scale factor cannot be derived due to the limited number of gluon splitting events in the \pt regime relevant for this analysis. The DDB scale factor is instead constrained in situ via the observed $\PZ\to\bbbar$ yield in the DDB pass and fail regions and constitutes the dominant experimental uncertainty.  The extracted DDB scale factors and uncertainties are shown in Table \ref{tab:ddb}.

\begin{table}[!htp]
\centering
\topcaption{Post-fit DDB tagger selection efficiency scale factor and uncertainty for different data-taking periods.}
\begin{tabular}{lllll}
{} & Scale factor  & \multicolumn{2}{c}{Uncertainty} \\ \hline
Early 2016 & $0.88$ & $+0.19$ & $-0.15$\\
Late 2016 & $0.92$ & $+0.18$ & $-0.13$\\
2017  & $0.79$ & $+0.12$ & $-0.10$\\
2018 & $0.85$ & $+0.11$ & $-0.10$\\ 
\end{tabular}
\label{tab:ddb}
\end{table}

The uncertainties on the corrections for substructure selection, jet mass scale, and jet mass resolution detailed in Section~\ref{sec:vjets} are treated as constrained nuisance parameters. These uncertainties are correlated among the \PW, \PZ, and Higgs boson production processes, and are further constrained by the presence of \PW and \PZ boson resonances in the $\msd$ distribution. Additional systematic uncertainties are applied to the event yields of $\PW$+jets, $\PZ$+jets, $\ttbar$, and Higgs boson production to account for the uncertainties due to the jet energy scale and resolution~\cite{CMS:2016lmd,cmsJESJER} and the limited event count of simulated samples. A bin-by-bin uncertainty is applied to account for the limited number of simulated QCD events in the dedicated fit to determine the transfer factor $T_\text{P/F}$ .

Other experimental uncertainties, including those related to the determination of the integrated luminosity~\cite{cmsLumi16,CMS-PAS-LUM-17-004,CMS-PAS-LUM-18-002}, variations in the amount of pileup obtained by adjusting the total inelastic cross section, modeling of the trigger efficiency, and the isolation and identification of the veto leptons are also considered, and contribute only a few percent to the total uncertainty.

In addition to experimental uncertainties, theoretical uncertainties are included in the final fit to account for inaccuracy in the modeling of SM processes.

The dominant theoretical uncertainty is due to the renormalization and factorization scales chosen for the simulated \PH samples. These uncertainties are applied to all H production processes and are propagated to the total expected yield of the H signal according to the prescription recommended in Ref.~\cite{becker2021precise}. They amount  to approximately 20\% and 5\% in the ggF and VBF signal yields, respectively. Uncertainties in the event yields due to initial- and final-state radiation are also calculated for all H production processes by varying the renormalization scale and non-singular term using the PYTHIA 8 showering algorithm \cite{Mrenna_2016}.  The Hessian PDF uncertainty is also applied to \PH signal yield, according to the prescription of Ref.~\cite{Butterworth_2016}.

Theoretical uncertainties on the $\PW$+jets and $\PZ$+jets processes account for missing higher-order QCD and mixed QCD-EW effects beyond the corrections described in Section~\ref{sec:samples}, following the prescription of Ref.~\cite{Lindert_2017}.

All theoretical uncertainties are considered to be correlated across all data-taking periods, and experimental systematic uncertainties are considered to be uncorrelated, with the exception of a correlated component of the uncertainty on the integrated luminosity ($\mathcal{L}$). All systematic uncertainties are incorporated into the analysis via nuisance parameters and treated according to the frequentist paradigm~\cite{CMS-NOTE-2011-005}.

Among the largest sources of systematic uncertainty are the theoretical uncertainties in the Higgs boson production cross section, the size of simulated signal samples, and the uncertainty in the data-to-simulation scale factor on the DDB tagger selection.  Overall, the measurement is limited by the statistical uncertainty in data, including the limited sample size available to determine the QCD background yield under the Higgs boson peak via the transfer factor $T_\text{res}$, as discussed in Section~\ref{sec:qcdbkg}.

\section{Results}
\label{sec:results}

\begin{figure}[htbp]
  \centering
  \includegraphics[width=0.49\textwidth]{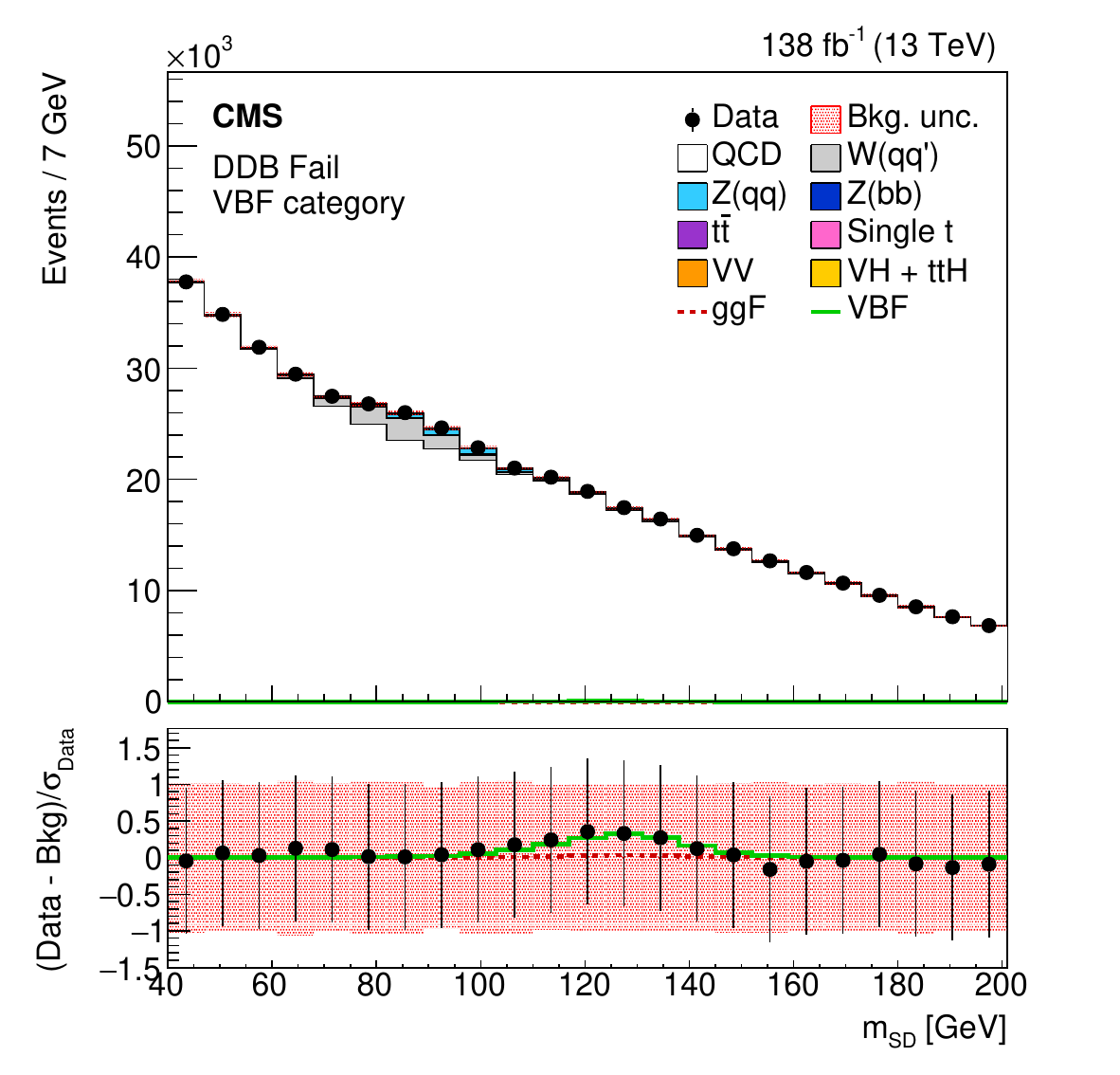}
  \includegraphics[width=0.49\textwidth]{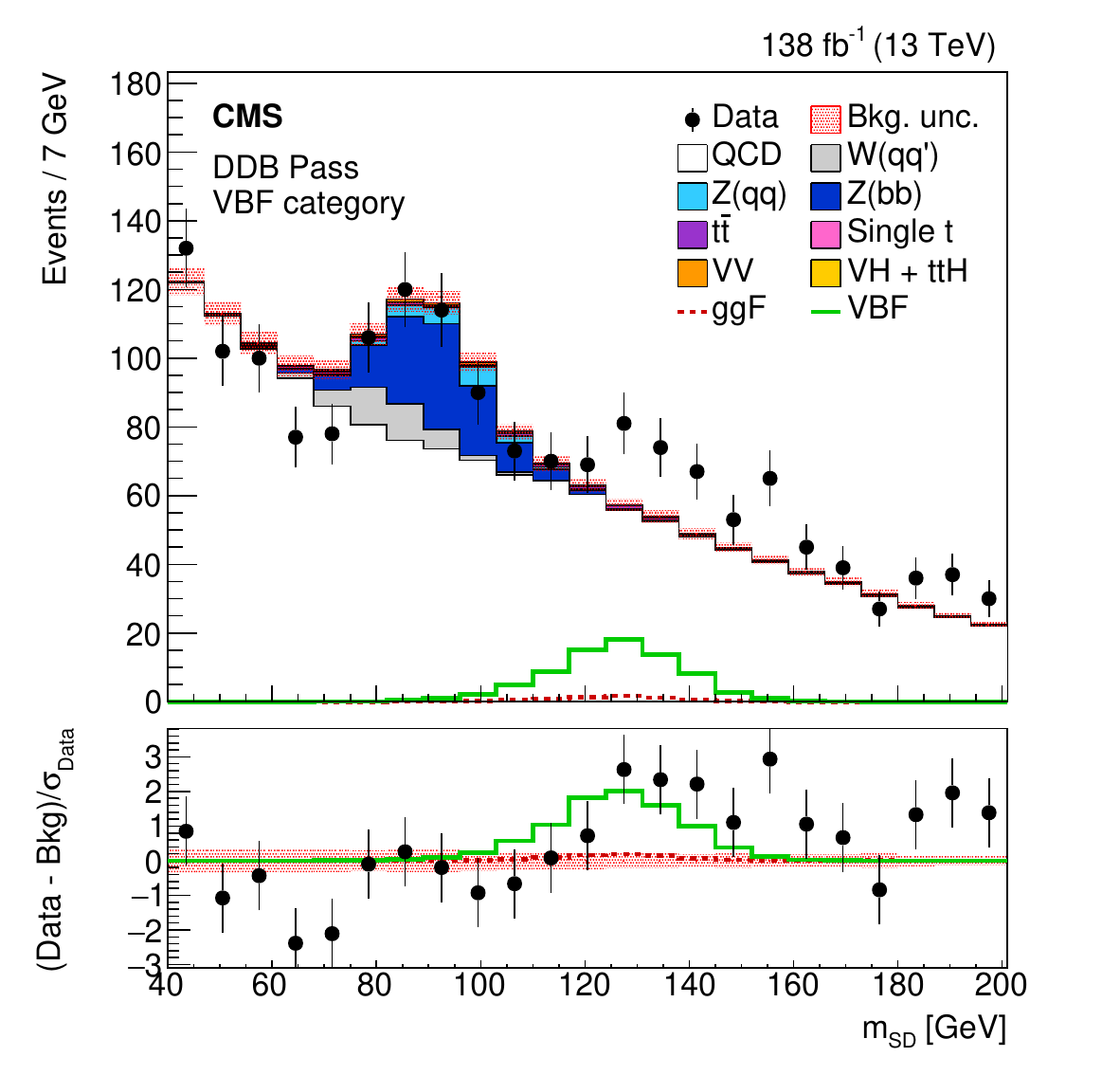}
  \caption{Post-fit soft drop mass distribution in the VBF category, summed over all $m_\text{jj}$ bins and data-taking periods for display purposes. The DDB fail (left) and pass (right) regions are shown.  The total background is broken down into contributions from different processes, and the total uncertainty is shown as a red band.  The lower panels show the difference between the data and background prediction divided by the statistical uncertainty in data. The near-perfect model agreement with data in the DDB fail region is by construction. The ggF and VBF distributions are overlaid in red and green, respectively. Each signal is scaled to its fitted event yield. }
  \label{result-vbf}
\end{figure}

\begin{figure}[htbp]
  \centering
  \includegraphics[width=0.49\textwidth]{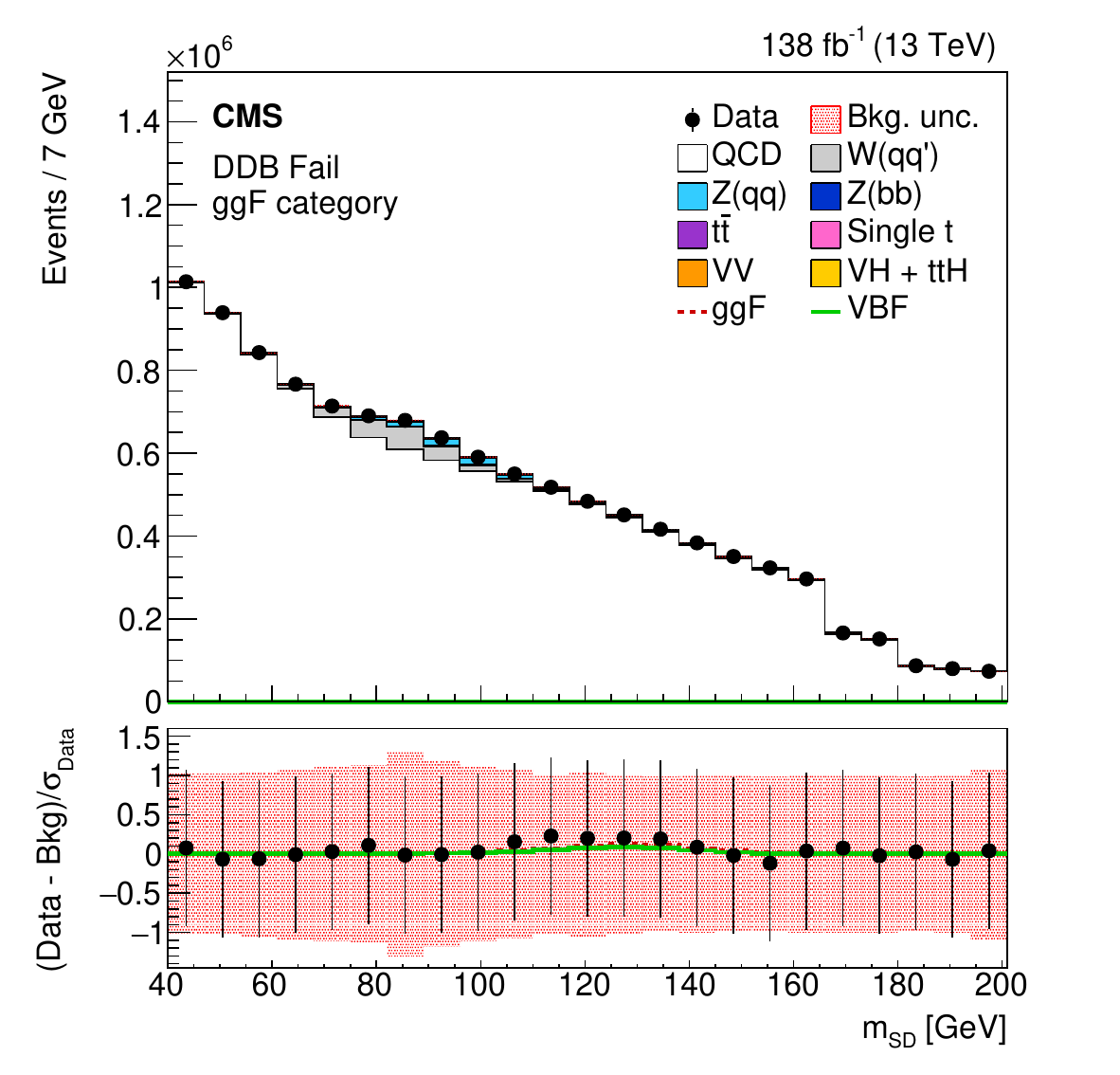}
  \includegraphics[width=0.49\textwidth]{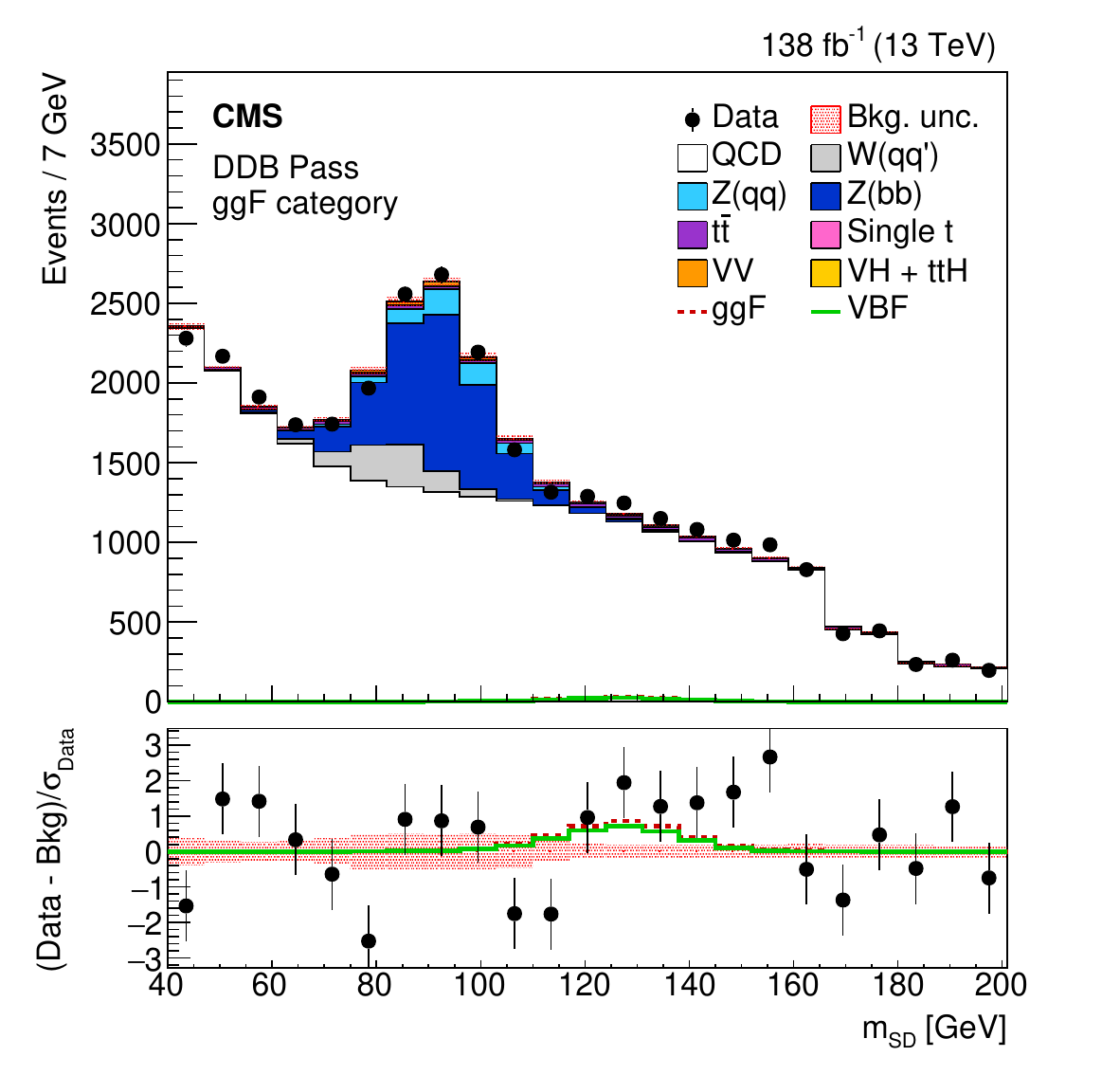} \caption{Post-fit soft drop mass distribution in the ggF category, summed over all $\pt$ bins and data-taking periods for display purposes. The DDB fail (left) and pass (right) regions are shown.  The total background is broken down into contributions from different processes, and the total uncertainty is shown as a red band.  The lower panels show the difference between the data and background prediction divided by the statistical uncertainty in data. The near-perfect model agreement with data in the DDB fail region is by construction. The ggF and VBF distributions are overlaid in red and green, respectively. Each signal is scaled to its fitted event yield. The apparent discontinuity at high mass is due to the exclusion of bins with extreme values of $\rho$.}
  \label{result-ggf}
\end{figure}

A binned maximum likelihood fit to the observed $\msd$ distributions is performed using the sum of the signal and background contributions. The test statistic chosen to determine the signal yield is based on the profile likelihood ratio~\cite{CMS-NOTE-2011-005}. The fit is performed simultaneously in many bins: in both $m_\text{jj}$ bins of the VBF category and all $\pt$ bins of ggF category, in the single-muon control region, in both the DDB pass and fail regions, and in all four data-taking periods. The observed data and post-fit $\msd$ distributions in the VBF category are shown in Fig.~\ref{result-vbf}. For display purposes, the data and fitted distributions are summed over both $m_\text{jj}$ bins and all four data-taking periods. For the ggF category, the observed data and post-fit $\msd$ distributions are shown in Fig.~\ref{result-ggf}, again summed over all six $\pt$ bins and all data-taking periods for display purposes. 

The best fit value of the signal strength, defined as the ratio of the measured to the SM expected cross section times $\PH\to\bbbar$ branching fraction, and an approximate 68\% confidence level (CL) interval are extracted following Ref.~\cite{Khachatryan_2015}. The fitted signal strengths for VBF and ggF \PH production, denoted $\mu_\text{VBF}$ and $\mu_\text{ggF}$, respectively, are extracted separately for the four data-taking periods, and a combined fit over the full data set is performed for the final result. The $\PV\PH$ and $\ttbar\PH$ processes are assumed to occur at SM rates.
The results are summarized in Table~\ref{tab-sig}.  The combined signal strength for the VBF process is measured to be $4.9^{+1.9}_{-1.6}$, and the combined signal strength for the ggF process is measured to be $1.6^{+1.7}_{-1.5}$. The expected signal strengths for VBF and ggF are $1.0_{-1.3}^{+1.4}$  and $1.0_{-1.3}^{+1.4}$, respectively. The observed (expected) correlation between these two signal strengths is $-0.41$ ($-0.41$). Tabulated results are provided in
the HEPData record for this analysis~\cite{hepdata}.

\begin{table}[]
  \renewcommand{\arraystretch}{1.3}
  \centering
  \topcaption{Fitted signal strength for $\PH\to\bbbar$ in the VBF and ggF channels for each data-taking period and for the full data set. The total uncertainty is shown, followed by the statistical component in parentheses.}
  \begin{tabular}{lrrrrr}
    {}         & $\mathcal{L}$ $[\fbinv]$ & VBF signal strength & ggF signal strength   \\ \hline
Early 2016 & 19.5 & $5.2^{+4.6}_{-3.8}$ ($^{+3.9}_{-3.6}$ stat)& $2.5^{+4.7}_{-4.3}$ ($^{+3.8}_{-3.7}$ stat)\\
Late 2016 & 16.8 & $5.6^{+5.8}_{-4.2}$ ($^{+4.5}_{-4.0}$ stat) & $0.6^{+4.4}_{-4.8}$ ($\pm3.8$ stat)\\
2017 & 41.5 & $0.8^{+2.8}_{-2.5}$ ($^{+2.7}_{-2.5}$ stat)& $3.3^{+3.1}_{-2.7}$ ($\pm2.5$ stat)\\
2018 & 59.8 & $8.3^{+3.9}_{-3.0}$ ($^{+2.8}_{-2.7}$ stat)& $0.4^{+2.6}_{-2.7}$ ($\pm2.3$ stat)\\ \hline
Combined & 138 & $4.9^{+1.9}_{-1.6}$ ($\pm1.5$ stat) & $1.6^{+1.7}_{-1.5}$ ($\pm1.4$ stat)\\
  \end{tabular}
  \label{tab-sig}
\end{table}

\begin{figure}[htbp]
  \centering
  \includegraphics[width=0.7\textwidth]{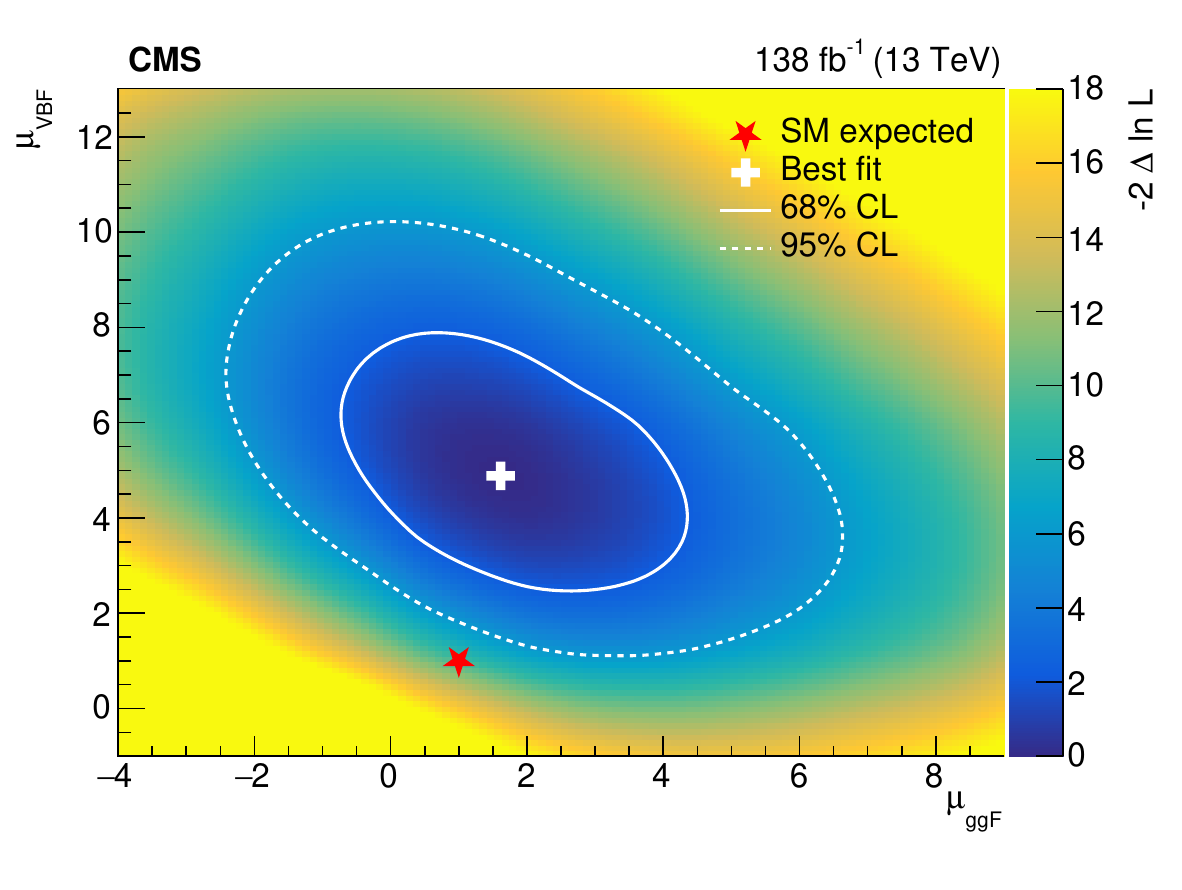}
  \caption{Two-dimensional likelihood contour of the VBF and ggF signal strengths. The color scale represents twice the negative log likelihood difference with respect to the best fit point. The observed 95\% (dashed) and 68\% (solid) contours are shown in white, and the best fit point as a white cross.  The SM expectation is marked by a red star.}
  \label{contour2d}
\end{figure}

Figure~\ref{contour2d} shows the combined two-dimensional likelihood scan performed over the VBF and ggF signal strengths.
The best fit point differs from the SM expectation by 2.7\,$\sigma$, and from the null hypothesis (no \PH production) by 4.0\,$\sigma$.

\subsection{Differential measurement}

\begin{figure}[htbp]
  \centering
  \includegraphics[width=0.49\textwidth]{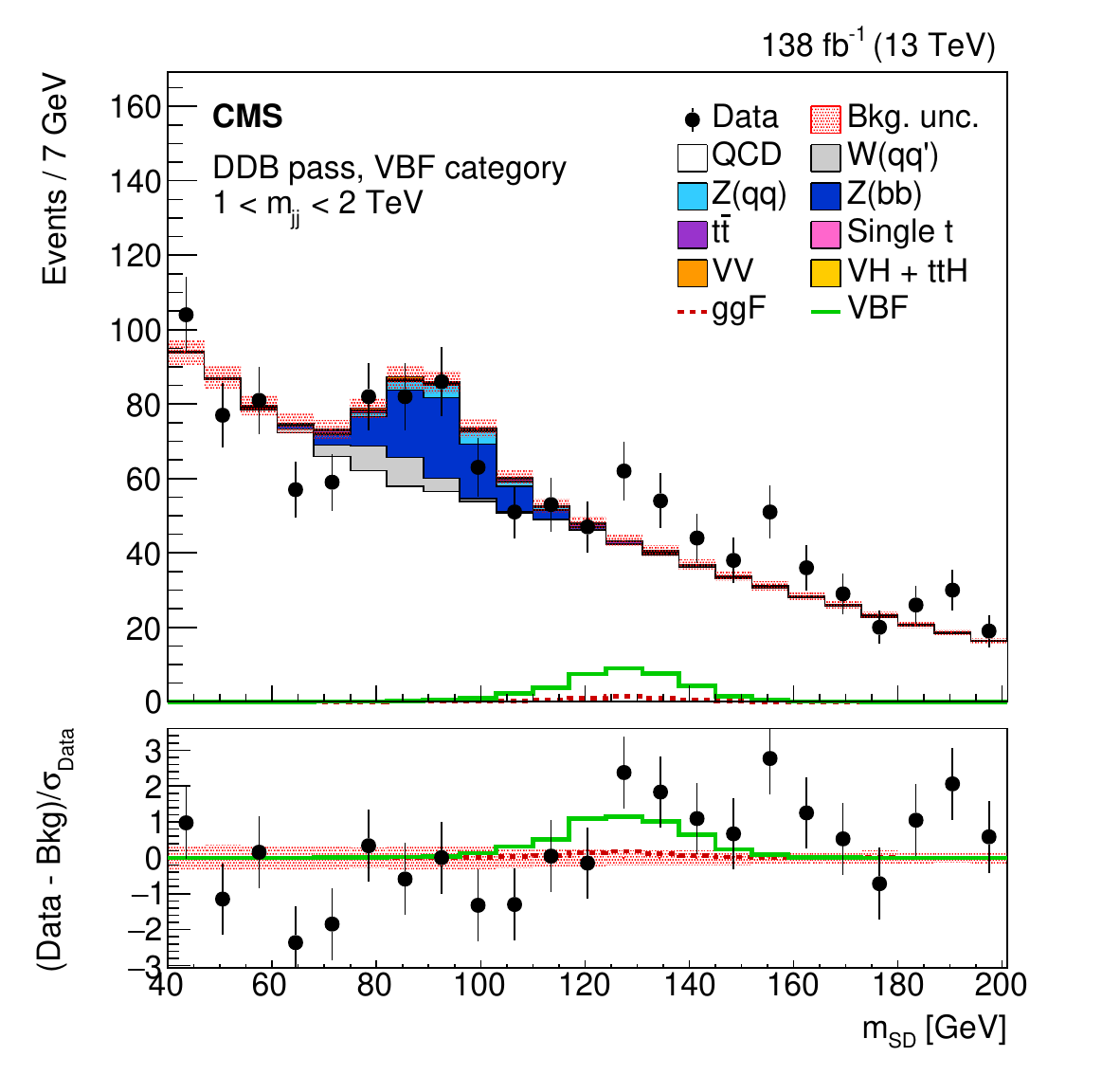}
  \includegraphics[width=0.49\textwidth]{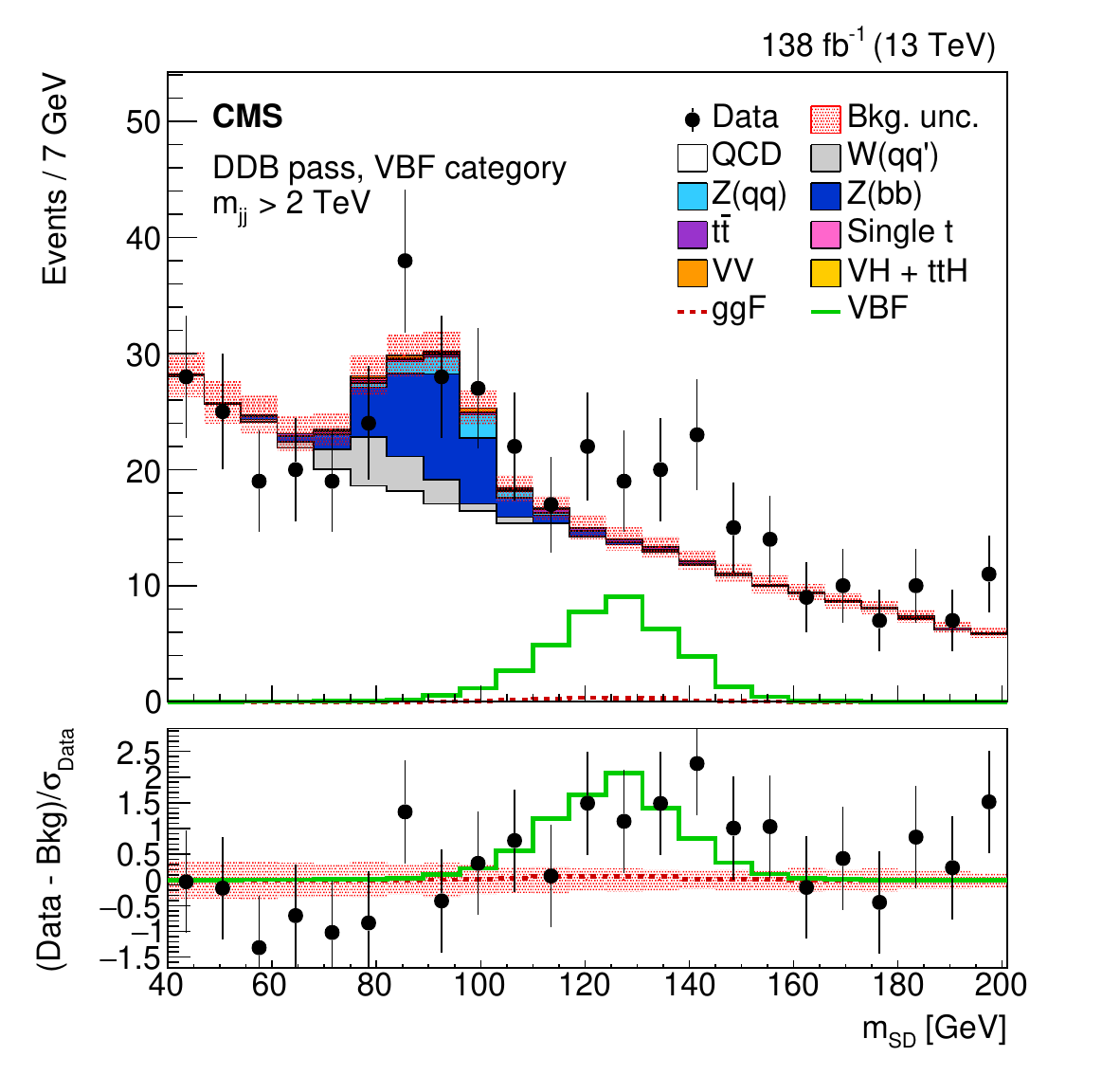}
  \caption{Post-fit soft drop mass distribution in each of the two $m_\text{jj}$ bins in the VBF category, summed over all data-taking periods for display purposes. The DDB pass region is shown. The lower panels show the difference between the data and background prediction divided by the statistical uncertainty in data. The ggF and VBF distributions are overlaid in red and green, respectively. Each signal is scaled to its fitted event yield. }
  \label{result-vbf-bins}
\end{figure}

\begin{figure}[htbp]
  \centering
  \includegraphics[width=0.49\textwidth]{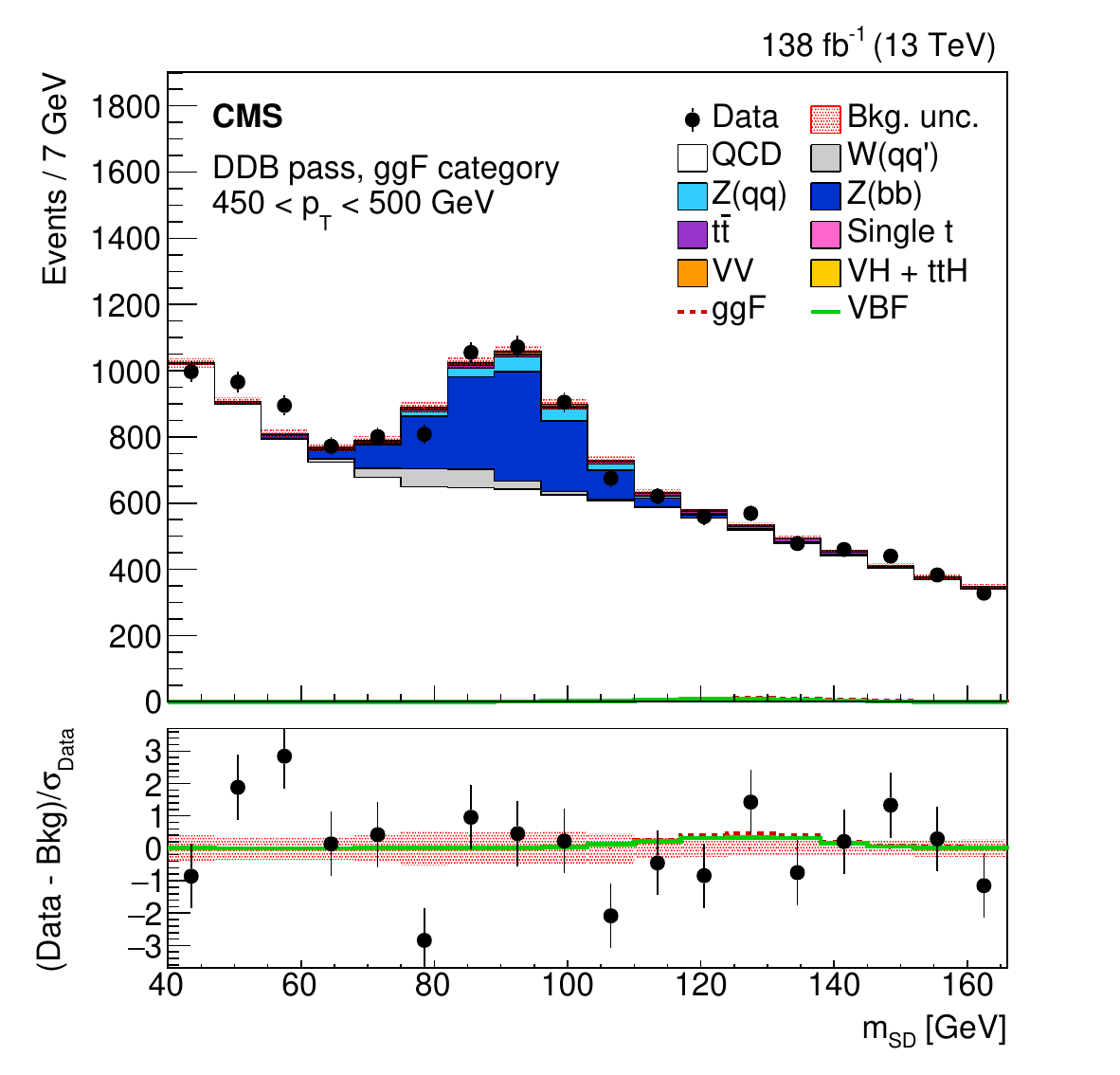}
  \includegraphics[width=0.49\textwidth]{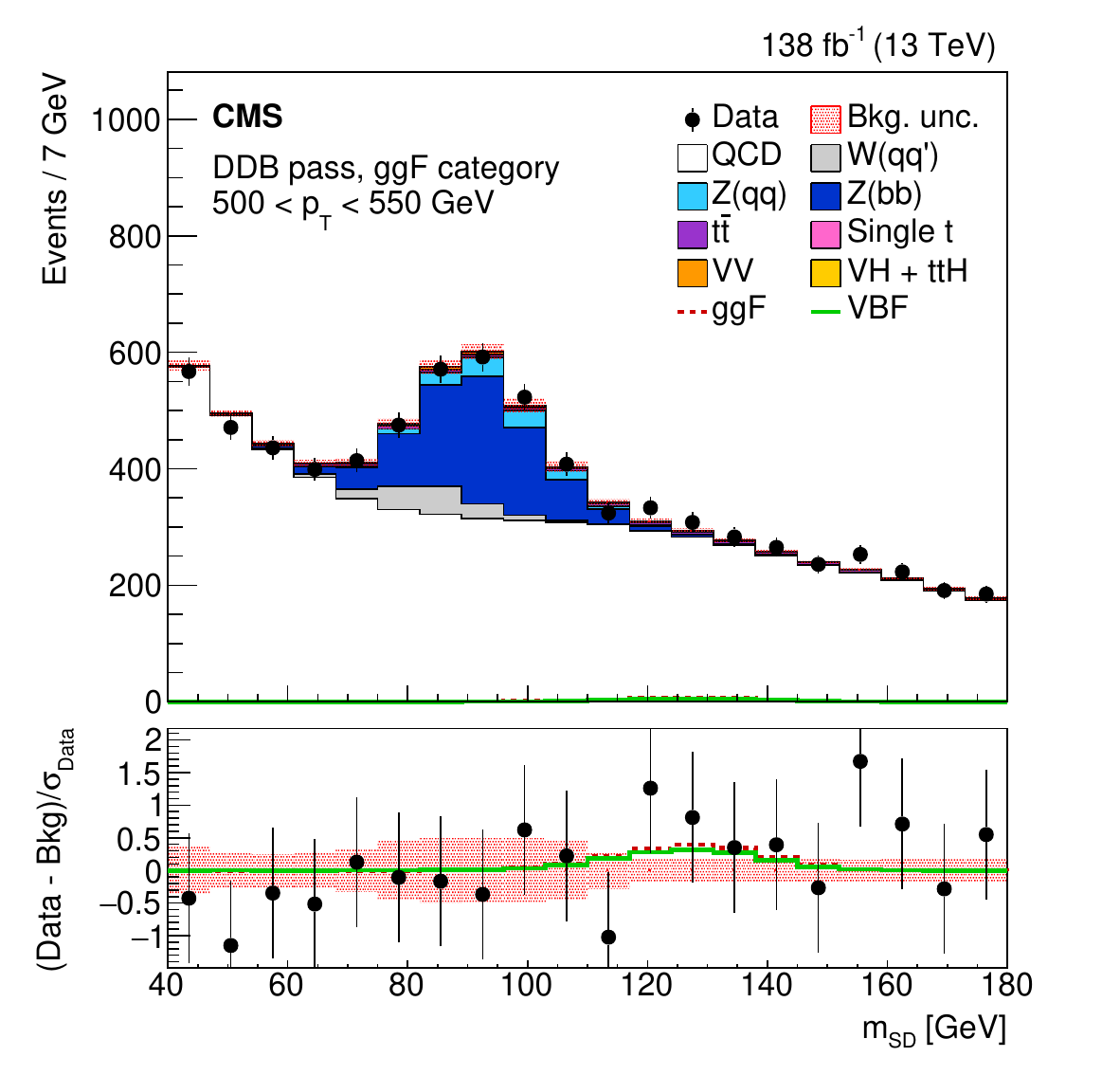} \\
  \includegraphics[width=0.49\textwidth]{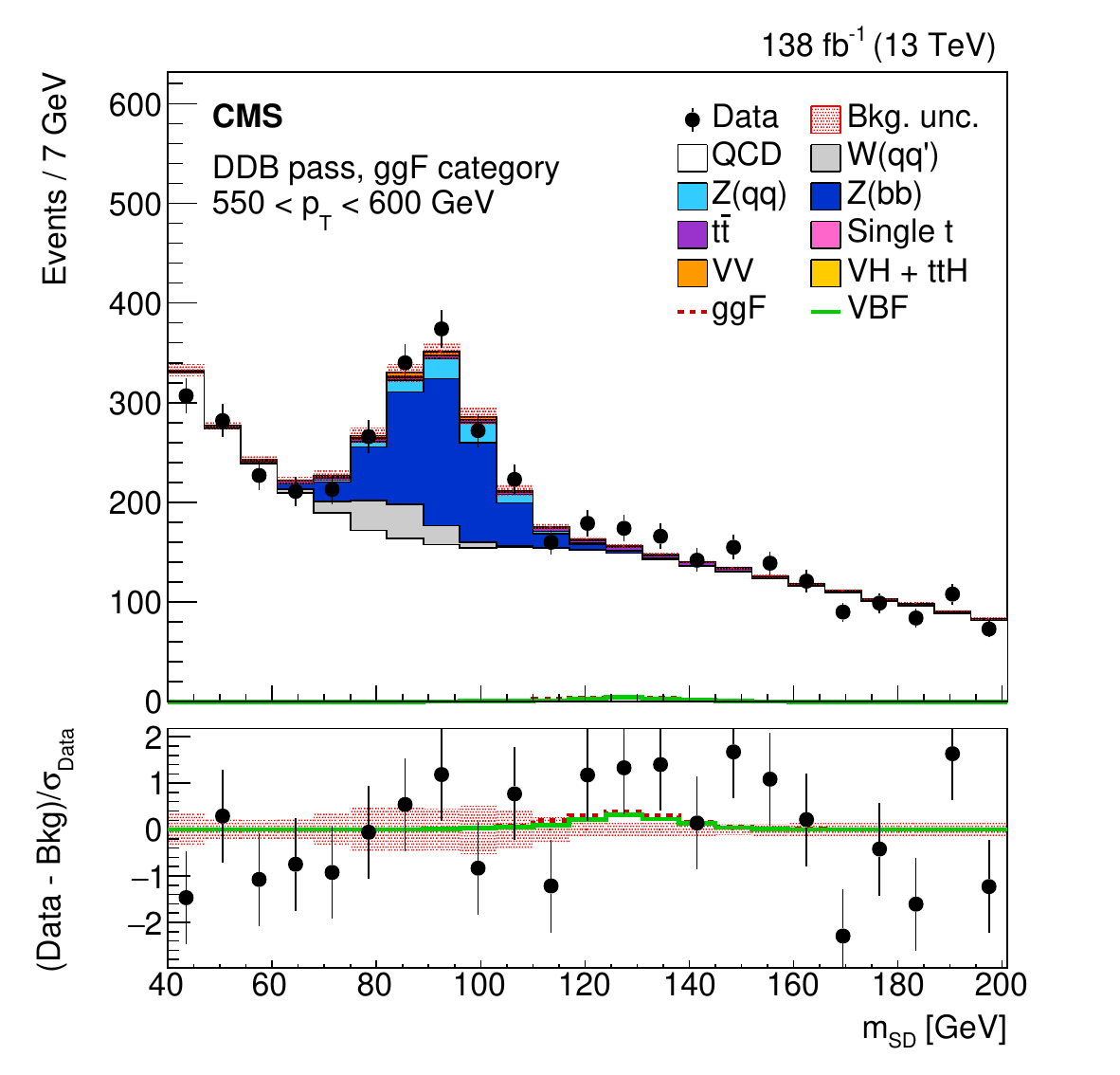}
  \includegraphics[width=0.49\textwidth]{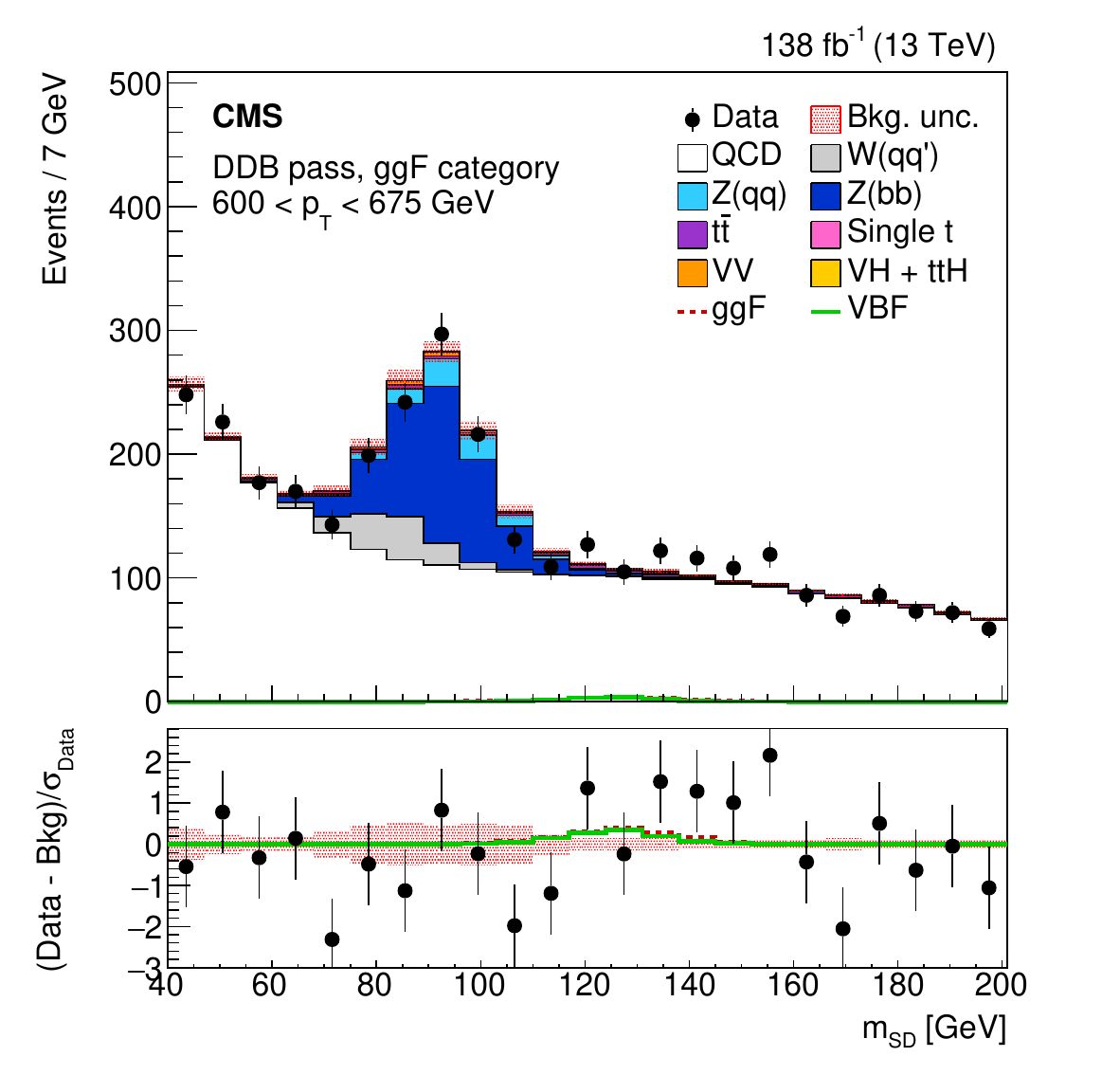} \\
  \includegraphics[width=0.49\textwidth]{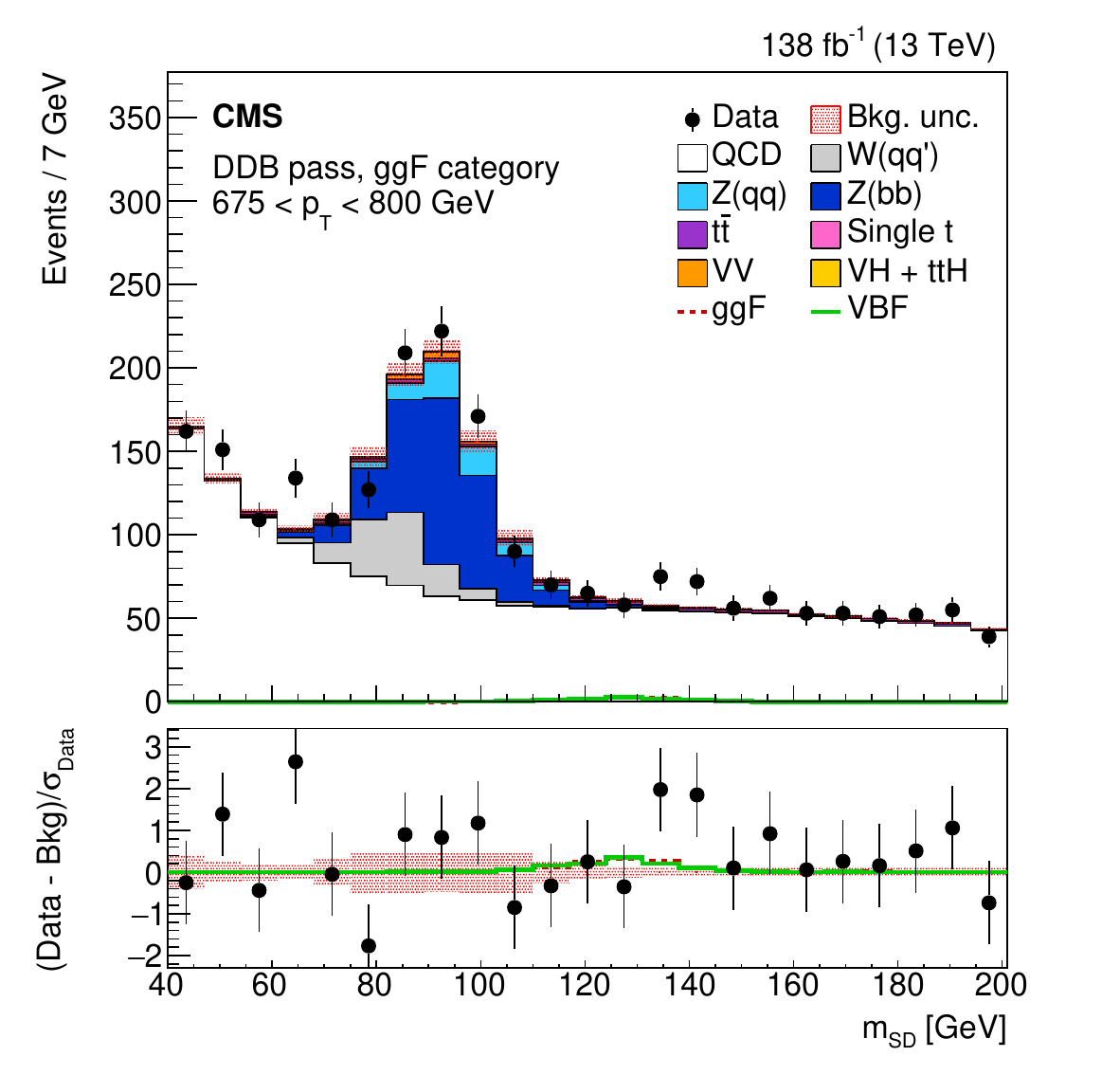}
  \includegraphics[width=0.49\textwidth]{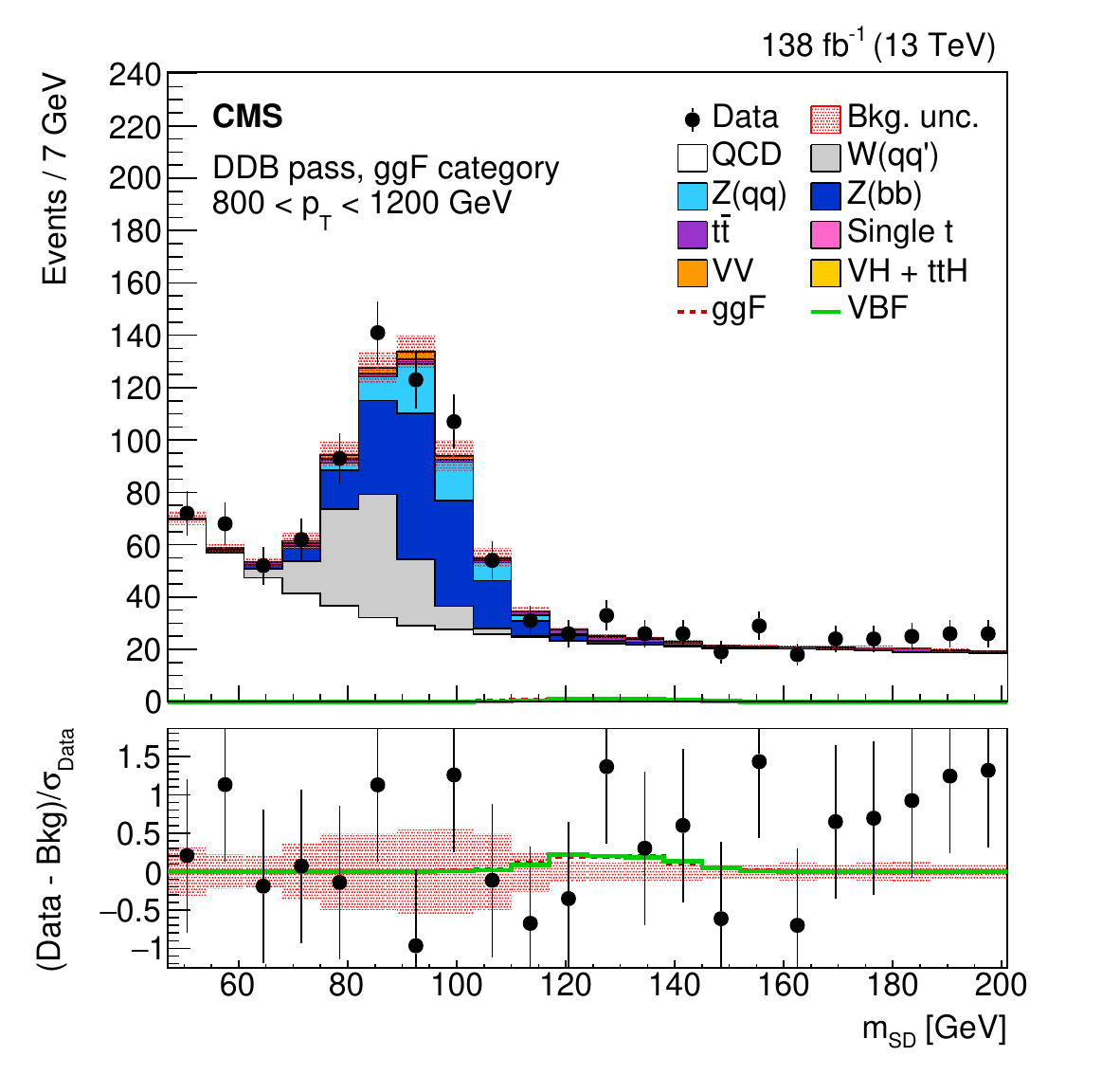}
  \caption{Post-fit soft drop mass distribution in each of the ${\pt}$ bins in the ggF category, summed over all data-taking periods for display purposes. The DDB pass region is shown. The lower panels show the difference between the data and background prediction divided by the statistical uncertainty in data. The ggF and VBF distributions are overlaid in red and green, respectively. Each signal is scaled to its fitted event yield. }
  \label{result-ggf-bins}
\end{figure}

\begin{figure}[htbp]
  \centering
  \includegraphics[width=0.8\textwidth]{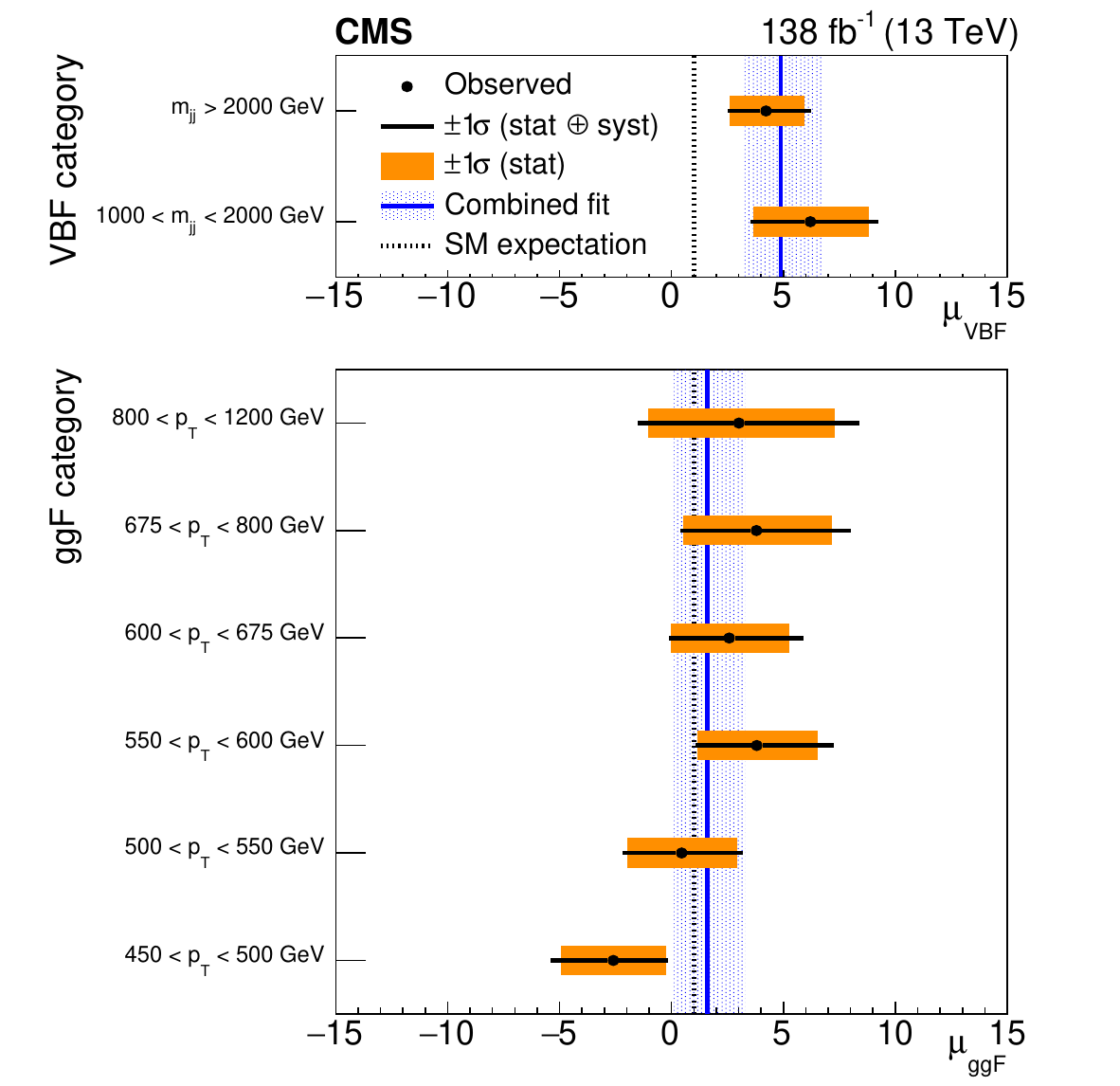}
  \caption{Upper: the VBF signal strength is shown in black, fitted per $m_\text{jj}$ bin, with the ratio of the ggF and VBF cross sections fixed to the value obtained in the signal strength fit. The horizontal black line represents the total uncertainty, and the orange bar represents the statistical-only component. The combined VBF signal strength and its uncertainty are shown in blue. The SM expectation is shown as a dashed line. Lower: the ggF signal strength is shown in black, fitted per $\pt$ bin, with the ratio of the ggF and VBF cross sections fixed to the value obtained in the signal strength fit. The horizontal black line represents the total uncertainty, and the orange bar represents the statistical-only component. The combined ggF signal strength and its uncertainty are shown in blue. The SM expectation is shown as a dashed line. }
  \label{result-bins}
\end{figure}

The observed $\msd$ distributions in the DDB pass region are shown separately in each bin in Figs.~\ref{result-vbf-bins} and~\ref{result-ggf-bins}, for VBF and ggF, respectively. The signal strength is also fitted independently per reconstruction-level bin. Because the VBF (ggF) category can only weakly constrain the ggF (VBF) process, the ratio of ggF to VBF is fixed to the value obtained in the signal strength fit described above.  The results are shown in Fig.~\ref{result-bins}.

A maximum likelihood unfolding technique~\cite{unfolding} is used to remove the effects of limited detector acceptance and response on the measured production cross section. The fiducial cross section is extracted simultaneously in five bins, each of which is considered as a separate process and modified by a freely floating signal strength parameter in the likelihood model. The three bins in the ggF channel correspond to fiducial regions with generated Higgs boson rapidity $|y^{\PH\text{,gen}}|<2.5$ and \pt ($\pt^{\PH,\text{gen}}$) in the ranges 300--450, 450--650, and $>650\GeV$. The two bins in the VBF channel correspond to rapidity $|y^{\PH\text{,gen}}|<2.5$, $\pt^{\PH,\text{gen}}>200\GeV$, and $m_\text{jj}^\text{gen}$ in the ranges 1000--1500\GeV and $>1500\GeV$, where $m_\text{jj}^\text{gen}$ is the generator-level invariant mass of the forward quark-initiated jets. This choice of bins is consistent with the simplified template cross section (STXS) stage 1.2 scheme ~\cite{deFlorian:2016spz,Badger:2016bpw}. Additional STXS bins corresponding to ggF with $\pt^{\PH,\text{gen}}<300\GeV$ and VBF with $\pt^{\PH,\text{gen}}<200\GeV$ or $m_\text{jj}^\text{gen}<1000\GeV$ are found to have negligible contribution to the signal region. The \PH signal yields in these bins are assumed to occur at SM rates.

For the theoretical uncertainties, only those that affect the acceptance of signal events in the analysis selection are taken into account.  Each fitted signal strength parameter and its uncertainty are scaled by the corresponding simulated cross section to obtain the best fit estimator for the cross section and its uncertainty.

\begin{figure}[htbp]
  \centering
  \includegraphics[width=0.8\textwidth]{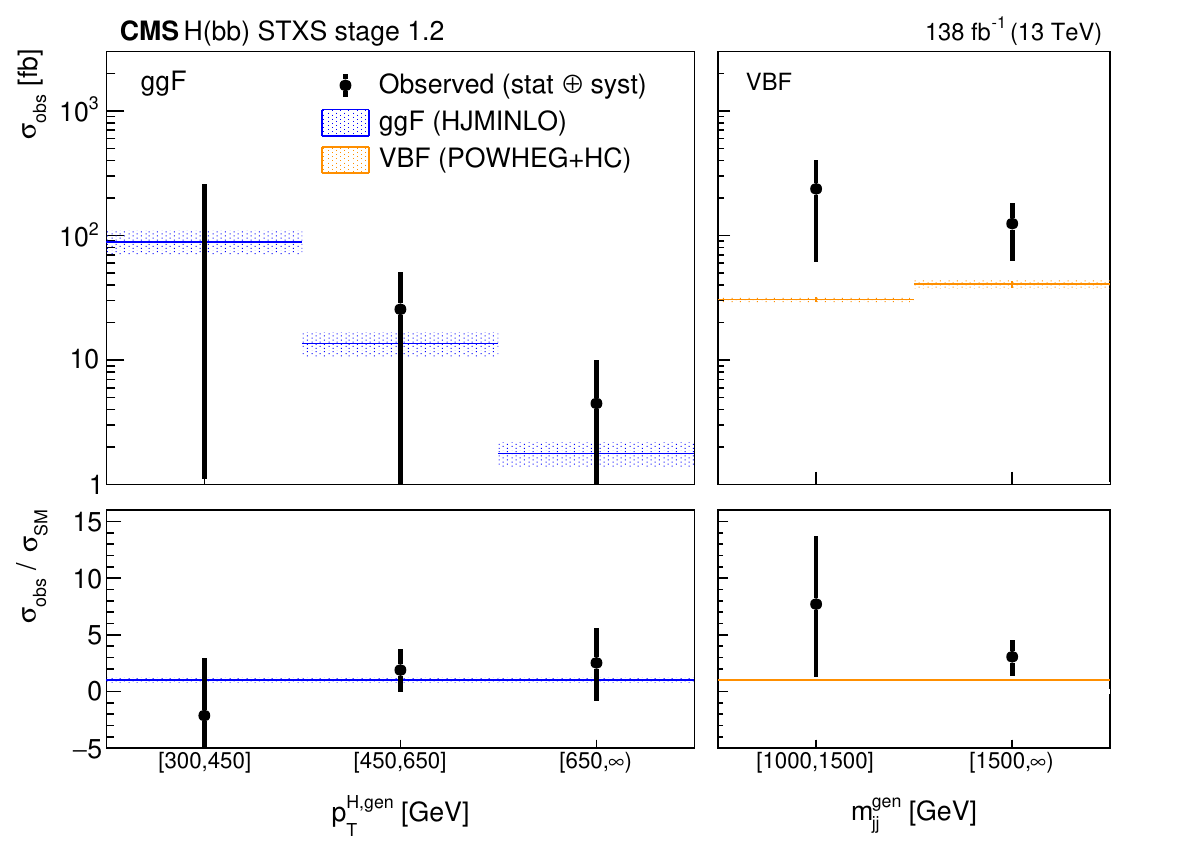}
  \caption{Unfolded cross section in five of the STXS stage 1.2 bins: three bins of Higgs boson \pt in ggF, and two bins of generator-level $m_\text{jj}$ in VBF. The SM prediction from HJMINLO and \POWHEG (plus higher order EW and NNLO QCD corrections) is overlaid for the ggF and VBF production mechanisms, respectively. The lower panel shows the ratio of the measured cross section to the SM prediction. }
  \label{unfolded}
\end{figure}

The unfolded fiducial cross sections are shown in Fig.~\ref{unfolded}, with the predictions from the signal event generators overlaid. The measured cross sections and uncertainties are also reported in Table~\ref{tab-unfolded}.
The best fit for the ggF bin with $300 < \pt^{\PH,\text{gen}} < 450\GeV$ lies below zero cross section, as can happen when the signal is estimated with a fit over a large background. However, the uncertainty in this bin is found to be comparatively large due to the limited number of these events passing the analysis selection, and the SM cross section value is well within the uncertainty.

\begin{table}[!htp]
\renewcommand{\arraystretch}{1.3}
  \centering
  \topcaption{Summary of the best fit estimators for the cross sections. In addition to the listed selection, each fiducial region has a rapdity requirement of $|y^{\PH\text{,gen}}|<2.5$. The third column shows the SM predicted cross section and uncertainty. The fourth column shows the fitted cross section, and the final column shows the total uncertainty followed by the statistical component in parentheses.}
  \begin{tabular}{llrrrr}
    Process & Fiducial region                                               & $\sigma_\text{SM}$ [fb] & $\sigma_\text{obs}$ [fb] & \multicolumn{2}{l}{Uncertainty [fb]}                          \\ \hline
ggF & $300 < \pt^{\PH,\text{gen}} < 450$ GeV & $90\pm20$ & $-190$ & $_{-450}^{+430}$ &($\pm410$ stat)\\
ggF & $450 < \pt^{\PH,\text{gen}} < 650$ GeV & $14\pm3$ & 26 & $_{-27}^{+25}$ & ($_{-23}^{+24}$ stat)\\
ggF & $\pt^{\PH,\text{gen}} > 650$ GeV & $2\pm0.4$ & $4$ & $_{-5}^{+6}$ & ($\pm5$ stat)\\ \hline
VBF & $\pt^{\PH,\text{gen}} > 200$ GeV,& \multirow{2}*{$31^{+1}_{-2}$} & \multirow{2}*{240} &  \multirow{2}*{$_{-190}^{+200}$} & \multirow{2}*{($\pm170$ stat)}\\
 & $1.0 < m_\text{jj}^{\text{gen}} < 1.5$ TeV \\
VBF & $\pt^{\PH,\text{gen}} > 200$ GeV,  & \multirow{2}*{$41^{+2}_{-3}$} & \multirow{2}*{120} & \multirow{2}*{$_{-61}^{+68}$} & \multirow{2}*{($_{-59}^{+62}$ stat)}\\
& $m_\text{jj}^\text{gen} > 1.5$ TeV \\ \hline
  \end{tabular}
  \label{tab-unfolded}
\end{table}
\clearpage

\section{Summary}

A measurement has been performed of boosted Higgs bosons (\PH) produced via vector boson fusion (VBF) or gluon fusion (ggF) and decaying to bottom quark-antiquark pairs. The analysis goes beyond the inclusive $\PH\to\bbbar$ measurements performed thus far to provide the first exploration of Higgs bosons produced with high transverse momentum ($\pt>450\GeV$) in the VBF channel.
The signal strengths for both processes are extracted simultaneously by performing a maximum likelihood fit to data in the large-radius jet mass distribution.  The observed signal strengths for the VBF and ggF processes are $4.9^{+1.9}_{-1.6}$ and $1.6^{+1.7}_{-1.5}$, corresponding to a 2.7\,$\sigma$ difference between data and the standard model expectation. The unfolded simplified template cross sections, which will provide an important input to future combined interpretations of \PH interactions, are also reported.

\begin{acknowledgments}
  We congratulate our colleagues in the CERN accelerator departments for the excellent performance of the LHC and thank the technical and administrative staffs at CERN and at other CMS institutes for their contributions to the success of the CMS effort. In addition, we gratefully acknowledge the computing centers and personnel of the Worldwide LHC Computing Grid and other centers for delivering so effectively the computing infrastructure essential to our analyses. Finally, we acknowledge the enduring support for the construction and operation of the LHC, the CMS detector, and the supporting computing infrastructure provided by the following funding agencies: SC (Armenia), BMBWF and FWF (Austria); FNRS and FWO (Belgium); CNPq, CAPES, FAPERJ, FAPERGS, and FAPESP (Brazil); MES and BNSF (Bulgaria); CERN; CAS, MoST, and NSFC (China); MINCIENCIAS (Colombia); MSES and CSF (Croatia); RIF (Cyprus); SENESCYT (Ecuador); ERC PRG, RVTT3 and MoER TK202 (Estonia); Academy of Finland, MEC, and HIP (Finland); CEA and CNRS/IN2P3 (France); SRNSF (Georgia); BMBF, DFG, and HGF (Germany); GSRI (Greece); NKFIH (Hungary); DAE and DST (India); IPM (Iran); SFI (Ireland); INFN (Italy); MSIP and NRF (Republic of Korea); MES (Latvia); LMTLT (Lithuania); MOE and UM (Malaysia); BUAP, CINVESTAV, CONACYT, LNS, SEP, and UASLP-FAI (Mexico); MOS (Montenegro); MBIE (New Zealand); PAEC (Pakistan); MES and NSC (Poland); FCT (Portugal); MESTD (Serbia); MCIN/AEI and PCTI (Spain); MOSTR (Sri Lanka); Swiss Funding Agencies (Switzerland); MST (Taipei); MHESI and NSTDA (Thailand); TUBITAK and TENMAK (Turkey); NASU (Ukraine); STFC (United Kingdom); DOE and NSF (USA).
    
  \hyphenation{Rachada-pisek} Individuals have received support from the Marie-Curie program and the European Research Council and Horizon 2020 Grant, contract Nos.\ 675440, 724704, 752730, 758316, 765710, 824093, 101115353, 101002207, and COST Action CA16108 (European Union); the Leventis Foundation; the Alfred P.\ Sloan Foundation; the Alexander von Humboldt Foundation; the Science Committee, project no. 22rl-037 (Armenia); the Belgian Federal Science Policy Office; the Fonds pour la Formation \`a la Recherche dans l'Industrie et dans l'Agriculture (FRIA-Belgium); the F.R.S.-FNRS and FWO (Belgium) under the ``Excellence of Science -- EOS" -- be.h project n.\ 30820817; the Beijing Municipal Science \& Technology Commission, No. Z191100007219010 and Fundamental Research Funds for the Central Universities (China); the Ministry of Education, Youth and Sports (MEYS) of the Czech Republic; the Shota Rustaveli National Science Foundation, grant FR-22-985 (Georgia); the Deutsche Forschungsgemeinschaft (DFG), among others, under Germany's Excellence Strategy -- EXC 2121 ``Quantum Universe" -- 390833306, and under project number 400140256 - GRK2497; the Hellenic Foundation for Research and Innovation (HFRI), Project Number 2288 (Greece); the Hungarian Academy of Sciences, the New National Excellence Program - \'UNKP, the NKFIH research grants K 131991, K 133046, K 138136, K 143460, K 143477, K 146913, K 146914, K 147048, 2020-2.2.1-ED-2021-00181, and TKP2021-NKTA-64 (Hungary); the Council of Science and Industrial Research, India; ICSC -- National Research Center for High Performance Computing, Big Data and Quantum Computing and FAIR -- Future Artificial Intelligence Research, funded by the NextGenerationEU program (Italy); the Latvian Council of Science; the Ministry of Education and Science, project no. 2022/WK/14, and the National Science Center, contracts Opus 2021/41/B/ST2/01369 and 2021/43/B/ST2/01552 (Poland); the Funda\c{c}\~ao para a Ci\^encia e a Tecnologia, grant CEECIND/01334/2018 (Portugal); the National Priorities Research Program by Qatar National Research Fund; MCIN/AEI/10.13039/501100011033, ERDF ``a way of making Europe", and the Programa Estatal de Fomento de la Investigaci{\'o}n Cient{\'i}fica y T{\'e}cnica de Excelencia Mar\'{\i}a de Maeztu, grant MDM-2017-0765 and Programa Severo Ochoa del Principado de Asturias (Spain); the Chulalongkorn Academic into Its 2nd Century Project Advancement Project, and the National Science, Research and Innovation Fund via the Program Management Unit for Human Resources \& Institutional Development, Research and Innovation, grant B39G670016 (Thailand); the Kavli Foundation; the Nvidia Corporation; the SuperMicro Corporation; the Welch Foundation, contract C-1845; and the Weston Havens Foundation (USA).
\end{acknowledgments}

\bibliography{auto_generated}

\cleardoublepage \appendix\section{The CMS Collaboration \label{app:collab}}\begin{sloppypar}\hyphenpenalty=5000\widowpenalty=500\clubpenalty=5000\input{HIG-21-020-public-authorlist.tex}\end{sloppypar}
\end{document}

%% file: HIG-21-020-public-authorlist.tex
\cmsinstitute{Yerevan Physics Institute, Yerevan, Armenia}
{\tolerance=6000
A.~Hayrapetyan, A.~Tumasyan\cmsAuthorMark{1}\cmsorcid{0009-0000-0684-6742}
\par}
\cmsinstitute{Institut f\"{u}r Hochenergiephysik, Vienna, Austria}
{\tolerance=6000
W.~Adam\cmsorcid{0000-0001-9099-4341}, J.W.~Andrejkovic, T.~Bergauer\cmsorcid{0000-0002-5786-0293}, S.~Chatterjee\cmsorcid{0000-0003-2660-0349}, K.~Damanakis\cmsorcid{0000-0001-5389-2872}, M.~Dragicevic\cmsorcid{0000-0003-1967-6783}, P.S.~Hussain\cmsorcid{0000-0002-4825-5278}, M.~Jeitler\cmsAuthorMark{2}\cmsorcid{0000-0002-5141-9560}, N.~Krammer\cmsorcid{0000-0002-0548-0985}, A.~Li\cmsorcid{0000-0002-4547-116X}, D.~Liko\cmsorcid{0000-0002-3380-473X}, I.~Mikulec\cmsorcid{0000-0003-0385-2746}, J.~Schieck\cmsAuthorMark{2}\cmsorcid{0000-0002-1058-8093}, R.~Sch\"{o}fbeck\cmsorcid{0000-0002-2332-8784}, D.~Schwarz\cmsorcid{0000-0002-3821-7331}, M.~Sonawane\cmsorcid{0000-0003-0510-7010}, S.~Templ\cmsorcid{0000-0003-3137-5692}, W.~Waltenberger\cmsorcid{0000-0002-6215-7228}, C.-E.~Wulz\cmsAuthorMark{2}\cmsorcid{0000-0001-9226-5812}
% ...
\par}
\vskip\cmsinstskip
\dag:~Deceased\\
$^{1}$Also at Yerevan State University, Yerevan, Armenia\\
$^{2}$Also at TU Wien, Vienna, Austria\\
% ...
